\documentclass[twocolumn]{autart}    % Enable this line and disable the 
\usepackage{amsmath,amssymb,amsfonts}
\usepackage{algorithm}
\usepackage{cite}
\usepackage[noend]{algpseudocode}
\usepackage{graphicx}
\usepackage{subfig}
\usepackage{caption}
\usepackage{hyperref}
\usepackage{enumerate}
\usepackage{color}
\usepackage{lipsum}
\newtheorem{definition}{Definition}

\newtheorem{theorem}{Theorem}
\newtheorem{corollary}{Corollary}
\newtheorem{assumption}{Assumption}

\newtheorem{proposition}{Proposition}
\newtheorem{remark}{Remark}
\newenvironment{proof}[1][Proof]
{\par\noindent\textbf{#1.}\ }
{\hspace*{\fill}$\square$\par}
\usepackage{mathrsfs} 
\begin{document}

\begin{frontmatter}

\title{Robust Bayesian Inference for Moving Horizon Estimation\thanksref{footnoteinfo}} % Title, preferably not more 

\thanks[footnoteinfo]{This study is supported by NSF China with U20A20334 and 52072213. It is also partially supported by Tsinghua University Initiative Scientific Research Program and Tsinghua University-Toyota Joint Research Center for AI Technology of Automated Vehicle. All correspondence should be sent to Shengbo Eben Li.}

\author[THU]{Wenhan Cao}\ead{cwh19@mails.tsinghua.edu.cn},    % Add the 
\author[PKU]{Chang Liu}\ead{changliucoe@pku.edu.cn},               % e-mail address 
\author[THU]{Zhiqian Lan}\ead{lanzq21@mails.tsinghua.edu.cn},  % (ead) as shown
\author[THU]{Shengbo Eben Li*}\ead{lisb04@gmail.com}
\author[UOM]{Wei Pan}\ead{wei.pan@manchester.ac.uk},
\author[ITA]{Angelo Alessandri}\ead{, angelo.alessandri@unige.it}

\address[THU]{School of Vehicle and Mobility, Tsinghua University, Beijing, China}  % Please supply                                              
\address[PKU]{Department of Advanced Manufacturing and Robotics, Peking University, Beijing, China}             % full addresses
\address[UOM]{Department of Computer Science, The University of Manchester, Manchester, UK}
\address[ITA]{University of Genova (DIME), Genova, Italy}

                             % keyword list or with the 
\begin{abstract} The accuracy of moving horizon estimation (MHE) suffers significantly in the presence of measurement outliers. Existing methods address this issue by treating measurements leading to large MHE cost function values as outliers, which are subsequently discarded. 
This strategy, achieved through solving combinatorial optimization problems, is confined to linear systems to guarantee computational tractability and stability. Contrasting these heuristic solutions, our work reexamines MHE from a Bayesian perspective, unveils the fundamental issue of its lack of robustness: MHE's sensitivity to outliers results from its reliance on the Kullback-Leibler (KL) divergence, where both outliers and inliers are equally considered. To tackle this problem, we propose a robust Bayesian inference framework for MHE, integrating a robust divergence measure to reduce the impact of outliers. 
In particular, the proposed approach prioritizes the fitting of uncontaminated data and lowers the weight of contaminated ones, instead of directly discarding all potentially contaminated measurements, which may lead to undesirable removal of uncontaminated data. A tuning parameter is incorporated into the framework to adjust the robustness degree to outliers. 
Notably, the classical MHE can be interpreted as a special case of the proposed approach as the parameter converges to zero. 
In addition, our method involves only minor modification to the classical MHE stage cost, thus avoiding the high computational complexity associated with previous outlier-robust methods and inherently suitable for nonlinear systems. Most importantly, our method provides robustness and stability guarantees, which are often missing in other outlier-robust Bayes filters.
The effectiveness of the proposed method is demonstrated on simulations subject to outliers following different distributions, as well as on physical experiment data.
\end{abstract}
\end{frontmatter}

\section{Introduction}
Accurate state estimation from noisy measurements, combined with the knowledge of system dynamics, is a crucial aspect of effective control systems. This problem has attracted significant attention across diverse domains, including vehicle control, aerospace navigation, and power grid management. Bayesian filtering offers a principled approach to solving state estimation problems by relying on Bayes' formula to update the posterior distribution of the system's state \cite{sarkka2013bayesian, chen2003bayesian}.
Building upon the foundation of Bayesian filtering, moving horizon estimation (MHE) has emerged as a powerful technique for obtaining accurate estimates in nonlinear systems \cite{rao2003constrained}. MHE is formulated as a receding horizon optimization problem, which takes into account a finite number of past measurements. This formulation enables MHE to handle diverse noise distributions and to account for constraints on states and noises. Indeed, MHE becomes the Kalman filter (KF) when constraints are absent and only the latest measurement is considered.

The accuracy of state estimation can be severely degraded when the assumed measurement generation mechanism deviates from the true mechanism. This misspecification of the measurement model, which may arise from sensor malfunction or inaccurate noise modeling, presents a significant challenge in maintaining estimation accuracy. Measurements affected by such misspecification, often referred to as outliers, necessitate careful handling to ensure the reliability of state estimates.

Outlier-robust MHE was first studied in \cite{alessandri2014moving,alessandri2016moving}, where a mixed-integer optimization problem was formulated to tackle the issue of measurement outliers. At each time instant, a set of least-squares cost functions is minimized, with the potential outlier-affected measurements being left out in turn. Subsequently, based on the assumption that outliers lead to a large value of cost function, only the minimizer associated with the lowest cost is selected as the estimate. A significant issue with this method is that simply discarding contaminated measurements can potentially lead to the loss of valuable information. Moreover, this strategy assumes bounded values of the measurement noise and intermittent outliers, which can limit its applicability as real-world noise and outliers often exhibit unbounded behavior and unpredictable distributions. 
Furthermore, it is constrained to be applicable only to linear systems in order to ensure stability and computational tractability, given the complexity of solving NP-hard mixed-integer optimization problems \cite{floudas1995nonlinear, alessandri2014moving,alessandri2016moving}.

To address the degradation of estimation accuracy caused by the presence of measurement outliers, robust Bayesian inference (RBI) provides a systematic solution \cite{knoblauch2022optimization,knoblauch2018doubly,akrami2022robust}. It is revealed in RBI theory that the sensitivity of Bayesian inference to outliers arises from its equal consideration of outliers and inliers. This sensitivity is due to the inherent mechanism of Bayesian inference, which aims to minimize the Kullback-Leibler (KL) divergence between the assumed measurement likelihood and the true data-generating mechanism. This revelation motivates us to reinterpret MHE from a Bayesian perspective, differing from the optimization viewpoint in most MHE literature, enabling the application of RBI to mitigate the impact of outliers rather than merely discarding contaminated measurements. However, directly applying RBI to Bayesian filtering lacks robustness and stability guarantees \cite{boustati2020generalised}, which are indispensable properties for state estimators. This theoretical gap highlights the necessity for a novel approach that can not only handle outliers effectively but also assure robustness and stability.

% Building on this insight, several outlier-robust Bayesian inference algorithms are developed\cite{knoblauch2022optimization,knoblauch2018doubly,akrami2022robust}. However, a critical challenge surfaces when applying RBI to Bayesian filtering: while it maintains meaningful probabilistic significance, it lacks the essential guarantees of stability and robustness, both of which are indispensable properties for state estimators \cite{boustati2020generalised}. This theoretical gap underscores the need for a strategy that can effectively manage outliers while concurrently ensuring robustness and stability.

Capitalizing on the theory of RBI and recognizing the theoretical gap when applying it to MHE, this paper proposes an outlier-robust MHE framework that maintains robustness and preserves stability in the presence of measurement outliers. We rectify the sensitivity issue of KL divergence by introducing the $\beta$-divergence to assign lower weight to outliers, thereby mitigating their impact. Furthermore, we present a thorough analysis of robustness and stability when applying RBI to MHE, compared to our prior work \cite{cao2023generalized}. Specifically, the key contributions of this paper can be summarized as follows:

\begin{enumerate}
\item We propose the $\beta$-MHE algorithm, which incorporates $\beta$-divergence into MHE for improved state estimation in the presence of outliers. This approach includes a tunable parameter, which allows for the adjustment of robustness degree against outliers. Notably, standard MHE is a special case of $\beta$-MHE when the parameter approaches zero. Further, the proposed algorithm necessitates only a minor modification to the MHE stage cost, circumventing the need for linearity assumptions and avoiding the NP-hardness issues associated with mixed-integer optimization in current robust MHE methods \cite{alessandri2014moving,alessandri2016moving}.

\item We analytically derive the influence functions for MHE methods, providing a solid foundation for analyzing their robustness.  Significantly, for linear Gaussian systems, we prove that the influence function of the proposed $\beta$-MHE is bounded, which demonstrates its robustness. This is the first work that systematically investigates the robustness of MHE from the perspective of influence function.

\item We analyze the stability of $\beta$-MHE by leveraging state-of-the-art stability results in MHE. We establish that, for a system exhibiting input/output-to-state stability and a cost function bounded by $\mathcal{K}$ functions with contraction properties, the proposed method ensures robustly asymptotically stable when the horizon length is sufficiently large.

\end{enumerate}
The remainder of this paper unfolds as follows: Section \ref{sec.related works} surveys the relevant literature. The problem formulation and preliminaries are presented in Section \ref{sec.III}. Our proposed algorithm for outlier-robust MHE is introduced in Section \ref{sec.IV}. Section \ref{sec.V} and Section \ref{sec.VI} study robustness and stability of $\beta$-MHE respectively. Simulation results are demonstrated in Section \ref{sec.VII}, while experimental results are provided in Section \ref{sec.VIII}. Finally, we wrap up with conclusions in Section \ref{sec.VIIII}.

\textbf{Notation:} $\mathcal{D}_{\text{KL}}(\cdot||\cdot)$ and $\mathcal{D}_{\beta}(\cdot||\cdot)$ are the Kullback–Leibler divergence and the $\beta$-divergence of two probability distributions, respectively. $\delta(\cdot,\cdot)$ represents the Dirac function. The set of real numbers is denoted as $\mathbb{R}$, and the set of nonnegative real numbers is denoted as $\mathbb{R}_{+}$. The set of integers that lie in the interval $[a, b]$ is denoted by $\mathbb{Z}_{[a, b]}$.
A function $\alpha: \mathbb{R}_{+}\to\mathbb{R}_{+}$ is a $\mathcal{K}$ function if it is continuous, strictly monotone increasing and $\alpha(0)=0$. 
% A function $\alpha: \mathbb{R}_{+}\to\mathbb{R}_{+}$ belongs to a $\mathcal{K}_{\infty}$ function if it is $\mathcal{K}$ function and $\alpha(x)\to\infty$ as $x\to\infty$. 
A function $\alpha: \mathbb{R}_{+}\to\mathbb{R}_{+}$ is a $\mathcal{L}$ function if it is continuous, nonincreasing, and $\alpha\to0$ if $t\to\infty$. A function $\alpha: \mathbb{R}_{+}\times\mathbb{R}_{+}\to\mathbb{R}_{+}$ is a $\mathcal{KL}$ function if for $\forall t$, $\alpha(\cdot, t)$ is a $\mathcal{K}$ function and $\forall s$, $\alpha(s, \cdot)$ is a $L$ function. 
$M \succeq 0$ indicates that the matrix $M$ is positive semi-definite, while $M \succ 0$ indicates that the matrix $M$ is positive-definite. $\| x \|$ represents the $l_2$-norm of the vector $x$, i,e., $\| x \| =\sqrt{x^{\top}x}$. 
We denote $\| x \|_{a:b} := \max_{t \in \mathbb{Z}_{[a, b]}}\left\{ \|x_t\| \right\}$ and $\| x \|_{0:\infty} := \sup_{t \in \mathbb{Z}_{[0, \infty]}}\left\{ \|x_t\| \right\}$.
For $M\succeq0$, $\Vert x \Vert_M$ is short for $\sqrt{x^{\top}Mx}$.  
$\Vert M \Vert_{\mathcal{F}}$ is the Frobenius norm of matrix $M$ while $|M|$ signifies the determinant of the matrix $M$. Besides, $\mathbb{I}_{n \times n}$ is the identity matrix with dimension $n$. 

\section{Related Works}\label{sec.related works}

% \wei{you actually pose several challenges, you should briefly state why you make the following classification of the subsections, and you should keep mentioned the CONs of existing method to give the reviewer an impression that the existing methods have their advantages, but can't solve all the challenges you posed.}
In this section, we first introduce existing Bayesian filtering techniques besides MHE, which typically ignore the presence of outliers. Subsequently, we examine how these traditional techniques have been modified to tackle the outlier problem. Finally, we broaden our scope to $H_{\infty}$ filtering, a robust state estimation method for outlier mitigation that operates outside the Bayesian filtering framework.
\subsection{Bayesian filtering}

Under the assumptions of linearity and Gaussianity, Bayesian filtering can simplify to the well-known KF, which acquires state estimates recursively and analytically \cite{kalman1960new}. However, a closed-form estimator like KF does not exist for nonlinear systems, as nonlinear functions cannot maintain the property of Gaussian. To circumvent this limitation, several KF variants have been proposed, such as the extended KF (EKF) and the unscented KF (UKF) \cite{julier2004unscented, ribeiro2004kalman}. Nevertheless, these methods still suffer from reduced estimation accuracy, particularly in highly nonlinear systems or under low signal-to-noise ratios. In contrast, particle filter (PF) uses
sequential Monte Carlo method to approximate the posterior distribution with a finite number of samples. While PF theoretically converges to the true distribution as the number of samples approaches infinity, its extensive computational requirements limit its practical use \cite{chen2003bayesian}. 
A significant weakness shared by all these methods is their sensitivity to measurement outliers. Such outliers can significantly impair estimation accuracy, emphasizing the need for robust estimation techniques.
% \wei{pro and con, especially con}

\subsection{Outlier-robust Bayesian filtering}

To address the significant challenge of measurement outliers, various methods have been developed. For linear systems where KF is prevalent, numerous outlier-robust KF variants have emerged. For example, leveraging Schweppe’s proposal and the Huber function, a batch-mode maximum likelihood-type KF is proposed in \cite{gandhi2009robust}. Alternatively, the maximum correntropy KF substitutes the minimum mean square error criterion with the more robust maximum correntropy criterion \cite{chen2017maximum}. Furthermore, a novel outlier-robust KF framework using a statistical similarity measure has been developed recently, illustrating that existing methods can be recovered or approximated by choosing different similarity functions \cite{huang2020novel}.
% \wei{logic to the next paragraph broken}
% \wei{pro and con, especially con}

While these methods can address the challenge of outliers in linear systems, capitalizing on the inherent structure of KF, they do not constitute a comprehensive, principled approach within the broader Bayesian framework. When these methods are extended to nonlinear systems, they often translate directly into outlier-robust EKF and UKF using linearization or unscented transformation techniques \cite{liu2017maximum, zhao2018robust, zhao2016robust}. However, this direct translation lacks a solid theoretical foundation. In contrast, outlier-robust PFs have been developed independently, without relying on the recursive form of KF. They utilize techniques such as kernel density estimation \cite{kumar2010new}, auxiliary particle filtering schemes \cite{maiz2009particle}, and hypothesis testing \cite{maiz2012particle} to bolster their robustness against outliers. Regrettably, these methods for nonlinear systems lack robustness guarantees and stability analyses compared to the robust filters designed for linear systems, such as in  \cite{gandhi2009robust,alessandri2014moving,alessandri2016moving}. This gap underlines the need for designing an outlier-robust filter for nonlinear systems with robustness and stability guarantees.

\subsection{\texorpdfstring{$H_{\infty}$}{H-infinity} filtering}
% \wei{again, logic broken with previous subsection..}\wei{pro and con, especially con}
In our discussion of robust filtering approaches, the $H_{\infty}$ filter merits attention due to its unique objectives and robustness characteristics. Unlike the Kalman filter, which aims to minimize the mean-square error, the $H_{\infty}$ filter is designed to constrain the worst-case variance in state estimates, ensuring the minimization of estimation energy error for all fixed-energy disturbances \cite{simon2006optimal}. Extensions for nonlinear systems such as the $H_{\infty}$ EKF, $H_{\infty}$ EKF and $H_{\infty}$ PF \cite{einicke1999robust,li2010h,wang2011h} have been developed, addressing nonlinear dynamics while maintaining robustness. However, the $H_{\infty}$ filtering approach is inherently conservative due to its design philosophy, which emphasizes worst-case scenarios. This results in a strategy that does not fully capitalize on available system model information, consequently leading to suboptimal performance.

\section{Problem Formulation and Preliminaries}
\label{sec.III}
% \wei{I expect some opening sentences here}
In this section, we present the problem formulation of MHE, comparing the conventional optimization-based viewpoint and a Bayesian perspective. We also introduce the state-space model formulation of MHE, laying the groundwork for a deeper exploration of its stability. Concluding this section, we revisit the optimization-centric interpretation of Bayesian inference, which forms the foundation for implementing RBI techniques.

\subsection{Problem Formulation in a Bayesian View}

MHE can be interpreted from two distinct viewpoints. Traditional research considers MHE as an optimization method, where a trajectory of state estimates is optimized online to fit the measurements with minimal noise estimates. In contrast, our perspective considers MHE as a tool for obtaining the maximum a posteriori (MAP) estimate within the framework of Bayesian inference.

Consider the stochastic systems characterized by the hidden Markov model (HMM): 
\begin{equation} \label{eq.hmm}
\begin{aligned}
x_0 &\sim \pi_0(x_0),\\
x_t|x_{t-1} &\sim f(x_t|x_{t-1}),\\
y_t|x_t &\sim g(y_t|x_t),
\end{aligned}
\end{equation}
where $x_t$ represents the unobserved states of a Markov process and $y_t$ denotes the associated noisy measurements. Besides, $\pi_0$ represents the initial distribution of the state, $f$ is the transition probability, and $g$ is the likelihood function, also known as the emission probability. In this context, optimal state estimation aims to infer the true state $x_t$ of stochastic systems based on current and past measurements $y_{1:t}$. To this effect, Bayesian filtering is utilized to compute either the posterior marginal distribution $q_t(x_t|y_{1:t})$ or the posterior joint distribution $q_t(x_{0:t}|y_{1:t})$ that is factorized as
\begin{equation} \label{eq.bf}
q_t(x_{0:t}|y_{1:t}) = \pi_0(x_0)\prod \limits_{i=1}^t g(y_i|x_i)f(x_i|x_{i-1}).
\end{equation}
For nonlinear systems with non-Gaussian noises, computing the posterior distributions becomes a challenging task. It is often more feasible to obtain a point estimate of the state instead of computing the entire density. For nonlinear systems, it is common to choose the MAP estimate due to the asymmetric and potentially multi-modal nature of the posterior distribution \cite{rawlings2006particle}. In light of this, full information estimation (FIE) targets the mode of the posterior joint distribution:
\begin{equation}\label{eq.fie}
\hat{x}_{0|t}, \hat{x}_{1|t},..., \hat{x}_{t|t} = \arg\max_{x_{0:t}}q_t(x_{0:t}|y_{1:t}).
\end{equation} 
However, the computational complexity of solving FIE in \eqref{eq.fie} rises with the number of available measurements. To maintain computational feasibility, MHE employs a sliding window approach, using a finite number of past measurements to simplify the problem. By leveraging logarithmic transformations, the MHE problem can be formulated as
%$= \hat{x}^*_{t-T:t}$
\begin{subequations}
\begin{align}
&\hat{x}_{t-T|t}, \hat{x}_{t-T+1|t}, ..., \hat{x}_{t|t} = \arg\min_{x_{t-T:t}}J(x_{t-T:t}),\label{eq.mhe-a}\\
&J(x_{t-T:t}) = \Gamma(x_{t-T}) +\sum\limits_{i=t-T+1}^{t} \left\{k(x_{i},x_{i-1})+h(y_i,x_i)\right\},\label{eq.mhe-b}
\end{align}
\end{subequations}
where 
\begin{subequations}
\begin{align}
\Gamma(x_{t-T})&\approx-\log{q_{t-T}(x_{t-T}|y_{1:t-T})}, \label{eq.mhe-1}\\
k(x_i,x_{i-1})&=-\log{f(x_{i}|x_{i-1})}, \label{eq.mhe-2}\\
h(y_i,x_i)&=-\log{g(y_{i}|x_{i})} \label{eq.mhe-3}.
\end{align}  
\end{subequations}
In this formulation, $\Gamma(x_{t-T})$ is referred to as the arrival cost (also known as the initial penalty), and $\mathbb{C}(x_t, x_{t-1}, y_t) = k(x_t,x_{t-1})+h(y_t,x_t)$ represents the stage cost function. For simplicity, we use $\hat{x}_{t-T:t}$ to represent $\left\{\hat{x}_{t-T|t}, \hat{x}_{t-T+1|t}, ..., \hat{x}_{t|t}\right\}$.
\begin{assumption}\label{assump.well-posed}
We assume that the optimization problem defined by \eqref{eq.mhe-a}\eqref{eq.mhe-b} is well-posed, i.e., the solution of \eqref{eq.mhe-a}\eqref{eq.mhe-b} exists for $\forall t \in \mathbb{Z}_{[0, \infty)}$.    
\end{assumption}
\begin{remark}
The arrival cost is a fundamental concept in MHE, serving as an equivalent statistic for summarizing past measurements $y_{1:t-T}$. Typically, when approximating the posterior distribution $q_{t-T}(x_{t-T}|y_{1:t-T})$ with a Gaussian distribution, the arrival cost is chosen to be of the quadratic form $\Vert x_{t-T}-\hat{x}_{t-T|t-T} \Vert^2_{P^{-1}_{t-T}}$. Here, $\hat{x}_{t-T|t-T}$ denotes the optimal estimate calculated in the previous time $t-T$, and $P_{t-T}$ is the error covariance matrix, which can be determined using EKF \cite{rao2003constrained} or UKF \cite{qu2009computation} through Ricatti equation. 
% \chang{This sentence is incomplete. Besides, should $\hat{x}_{t-T}$ use a different symbol to distinguish from $\hat{x}_{t-T:t}$? $P_{t-T}$ needs to be defined as well.}.    
\end{remark}

In practical scenarios, the assumption that measurements are strictly generated by the likelihood function $g$ often fails, leading to outliers that can critically undermine MHE's accuracy \cite{alessandri2014moving,alessandri2016moving}. Responding to this challenge, our work focuses on formulating an approach that can effectively handle these outliers within the MHE framework. Unlike previous outlier-robust MHE approaches that are constrained to linear systems and specific noise assumptions \cite{alessandri2014moving,alessandri2016moving}, our work takes into account more general nonlinear systems, eliminating the need for stringent noise assumptions. Furthermore, in contrast to other outlier-robust Bayesian filtering techniques \cite{liu2017maximum, zhao2018robust, zhao2016robust,kumar2010new,maiz2009particle, maiz2012particle} that lack robustness or stability guarantees, our methodology aims to provide these assurances, thereby improving the reliability of the estimator.

% MHE, as it is conventionally formulated, does not specifically address the presence of outliers in the measurement data. However, in real-world situations, the assumption that measurements are generated impeccably by the likelihood function $g$ is often not upheld \wei{citations are definitely needed here! maybe outlier is just a made-up problem}. Consequently, these imperfections in the data can cause issues in the estimation process. In light of this, the problem formulation in this paper focuses on addressing the issue of outliers within the MHE framework by developing a robust solution that specifically targets the mitigation of their effects, thus providing an effective approach to the problem at hand.

\subsection{State Space Model Formulation of MHE}
% \wei{can you glue the logic from section 3.1. I just feel too jumping}
While the HMM-based formulation of MHE provides a clear probabilistic interpretation, the state space model (SSM) formulation proves more convenient for stability analysis. SSM can be expressed as:
\begin{equation}\nonumber%\label{eq.state space model}
\begin{aligned}
x_{t+1} &= \mathcal{F}(x_t, \xi_t),\\
y_t &= \mathcal{G}(x_t)+ \zeta_{t}.
\end{aligned}
\end{equation}
Where $\mathcal{F}$ and $\mathcal{G}$ represent the transition and measurement functions, respectively, while $\xi_t$ and $\zeta_t$ denote the process and measurement noise. The transition function $\mathcal{F}$ in SSM aligns with the transition probability $f$ in HMM, with the measurement function $\mathcal{G}$ aligning with the likelihood function $g$. For example, if the measurement noise adheres to a normal distribution, i.e., $\zeta_t \sim \mathcal{N}(0, R)$, we can express the likelihood function as $g(y_t|x_t) = \frac{1}{\sqrt{(2\pi)^{n_{\zeta}} |R|}} \exp \left(-\frac{1}{2}(y_t - \mathcal{G}(x_t))^{\top} R^{-1}(y_t - \mathcal{G}(x_t)\right)$, where $n_{\zeta}$ is the dimension of $\zeta_t$. 

For stability analysis, we can rewrite the stage cost $\mathbb{C}(x_t, x_{t-1}, y_t)$ as $\mathbb{C}(\xi_{t-1}, \zeta_t)= k(\xi_{t-1})+h(\zeta_t)$ and the objective function $J(x_{0:t})$ as $J(x_{0},\xi_{0:t-1})$ within the SSM framework. This transformation modifies the optimization variables to $\xi_{t-1}$ and $\zeta_t$. For instance, with $\zeta_t \sim \mathcal{N}(0, R)$, we can represent $h(y_t,x_t)$ as $h(\zeta_t)=\frac{1}{2}\zeta_t^{\top} R^{-1}\zeta_t$, significantly streamlining the notation. 
% This assumption is illustrated as:
% \begin{assumption}\label{assump.well-posed}\cite{rawlings2012optimization, allan2019lyapunov}
% $\mathcal{F}(\cdot)$, $\mathcal{G}(\cdot)$ and $\mathbb{C}(\cdot, \cdot
% )$ are continuous functions. $\mathbb{C}(\xi_t, \zeta_t)$ adheres to the following inequalities,
% \begin{equation}\nonumber
% \mathbb{C}(\xi_t, \zeta_t) \geq \underline{\alpha}_{\xi}(\Vert \xi_t \Vert) + \underline{\alpha}_{\zeta}(\Vert \zeta_t \Vert),
% \end{equation}
% in which $\underline{\alpha}_{\xi}$, $\underline{\alpha}_{\zeta}$ are $\mathcal{K}_{\infty}$ functions.
% \end{assumption}
\begin{remark}
In our study, we primarily focus on state estimation. Therefore, any control inputs that are known beforehand are treated as fixed constants. This treatment simplifies the ensuing optimization and associated analyses without sacrificing any critical information. For the sake of clarity and conciseness, these control inputs are omitted from the problem formulation.
\end{remark}

\subsection{An Optimization-centric View on Bayesian Inference} \label{GBI}
% \wei{logic broken, why optimisation?}
In this subsection, we review the optimization-centric perspective of Bayesian inference, laying the foundation of RBI and paving the way for our proposed approach. Bayesian inference calculates the posterior distribution $p_\text{pos}(x|y)$ utilizing a prior probability $\pi(x)$ and a likelihood function $g(y|x)$,
\begin{equation}\nonumber
\begin{aligned}
p_\text{pos}(x|y)=\frac{\pi(x)g(y|x)}{p(y)}, \; p(y)=\int_{x}{\pi(x)g(y|x) \mathrm{d}x}.
\end{aligned}
\end{equation} 
From an optimization-centric perspective on Bayesian inference \cite{knoblauch2022optimization}, the posterior distribution can be formulated as a solution to the optimization problem:
\begin{equation}\label{eq.BI optimization view}
\begin{aligned}
&p_{\text{pos}} = \min_{\nu}\left\{ \int{-\log{g(y|x)}\nu(x)\mathrm{d}x}+\mathcal{D}_{\text{KL}}(\nu||\pi)\right\}, \\
&\mathcal{D}_{\text{KL}}(\nu||\pi)  := \int \nu(x) \log \left( \frac{\nu(x)}{\pi(x)} \right) \,\mathrm{d}x.
\end{aligned}
\end{equation}
Indeed, the negative logarithm of the likelihood in \eqref{eq.BI optimization view}, i.e., $-\log{g(y|x)}$, can be extended to a generalized loss function $\ell(x,y)$. This function captures the discrepancy between observed data and the assumed likelihood, resulting in a generalized optimization problem:
\begin{equation}\label{eq.gbi}
q = \min_{\nu}\left\{\int{\ell(x,y)\nu(x)\mathrm{d}x}+\mathcal{D}_{\text{KL}}(\nu||\pi)\right\}.
\end{equation}
The resulting belief distribution $q$, also referred to as the Gibbs posterior \cite{syring2017gibbs}, leads to a Bayes-type updating rule given by
\begin{equation}\nonumber
\begin{aligned}
q(x) = \pi(x)\frac{G(y|x)}{Z}, \; Z =\int_{x}{\pi(x)G(y|x) \mathrm{d}x},
\end{aligned}
\end{equation}
where $G(y|x):=\exp{\left(-\ell(x,y)\right)}$ denotes a generalized likelihood function. By strategically designing the loss function $\ell$, it is possible to achieve RBI, effectively mitigating the impact of outliers in the data.
\begin{remark}
The optimization target of \eqref{eq.BI optimization view} is essentially the negative of the evidence lower bound (ELBO) used in variational inference. In detail, variational inference aims to maximize the ELBO \cite{blei2017variational}, and our methodology seeks to minimize its negative counterpart.
\end{remark}

\section{Algorithm Design for Outlier-Robust MHE}\label{sec.IV}
In Section \ref{sec.III}, we showed that the negative logarithm of the likelihood in Bayesian inference could be extended to a generalized likelihood function $G(y|x)=\exp{\left(-\ell(x,y)\right)}$. This insight also inspires a generalization of Bayesian filtering. Specifically, if we define the sequence of generalized likelihoods as $G(y_t|x_t) = \exp{\left(-\ell(x_t,y_t)\right)}$, the joint generalized posterior distribution $q'_t(x_{0:t}|y_{1:t})$ can be expressed as the product of $G(y_t|x_t)$ and $f(x_t|x_{t-1})$, i.e.,
\begin{equation} \label{eq.gbf}
q'_t(x_{0:t}|y_{1:t}) = \pi_0(x_0)\prod \limits_{i=1}^t G(y_i|x_i)f(x_i|x_{i-1}).
\end{equation}

Observe that when the loss function $\ell(x_t,y_t)$ is chosen to be the negative logarithm of the likelihood, i.e., $\ell(x_t,y_t) = -\log{g(y_t|x_t)}$, \eqref{eq.gbf} becomes the posterior distribution in Bayesian filtering \eqref{eq.bf}. Under this circumstance, $G(y_t|x_t)=\exp{\left(\log{g(y_t|x_t)}\right)} = g(y_t|x_t)$.

Essentially, maximizing $-\log{g(y_t|x_t)}$ is equivalent to minimizing $\mathcal{D}_{\text{KL}}(p_\text{true}^t(y_t)||g(y_t|x_t))$ \cite{ghosh2016robust, futami2018variational}, which is the KL divergence between the true data-generating model, denoted as $p_\text{true}^t(y_t)$, and the assumed likelihood function $g(y_t|x_t)$. The sensitivity of KL divergence to outliers leads to the sensitivity of Bayesian filtering, as it treats all data points equally, without distinguishing between inliers and outliers \cite{akrami2022robust}. To enhance the robustness of Bayesian filtering, a more robust divergence such as the $\beta$-divergence can be employed \cite{ghosh2016robust, futami2018variational}. 

\begin{figure}[t]%[!htb]
\centering
\includegraphics[width=0.8\linewidth]{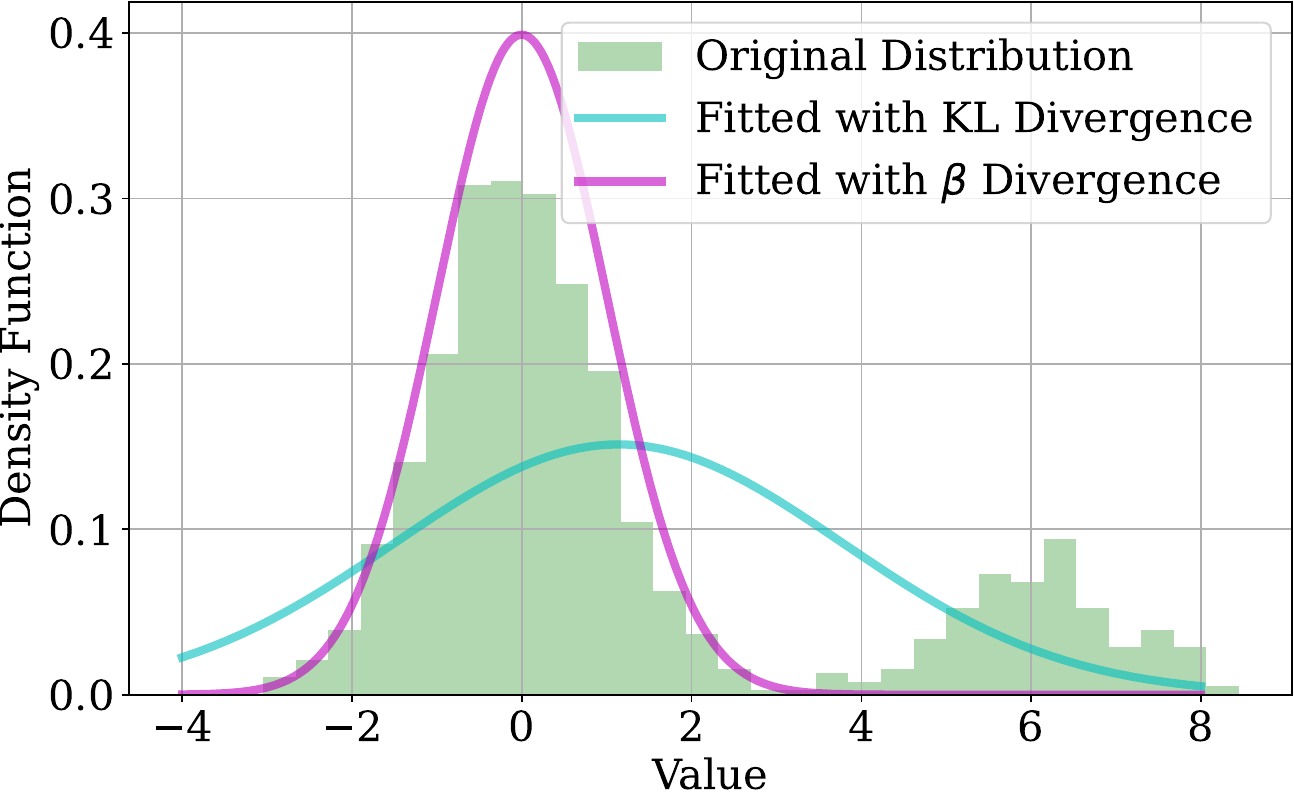}
\caption{Comparison of the robustness of $\beta$-divergence and KL divergence for handling outliers. Here, the Gaussian distribution $\mathcal{N}(0,\;1)$ is contaminated with Gaussian distribution $\mathcal{N}(5,\;1)$ with probability 0.1.}
\label{fig.influence}
\end{figure}

Fig. \ref{fig.influence} compares KL divergence and $\beta$-divergence for parameter estimation of a single Gaussian distribution. When optimizing KL divergence, both inliers and outliers are treated equally, leading to an estimate that can be significantly disturbed by outliers. Conversely, the optimization of $\beta$-divergence yields a more robust estimate. It de-emphasizes the impact of outliers and focuses on fitting the primary distribution \cite{futami2018variational}. This property makes $\beta$-divergence a suitable choice when dealing with measurement outliers.

The $\beta$-divergence between the true data distribution $p_\text{true}^t(\cdot)$ and the assumed likelihood $g(\cdot|x_t)$ is given by
\begin{equation}\label{eq.beta divergence}
\begin{aligned}
&\mathcal{D}_{\beta}\left(p_\text{true}^t(\cdot)||g(\cdot|x_t)\right)
=\frac{1}{\beta}\int{p_\text{true}^t(y)^{\beta+1}\mathrm{d}y} \\&- \frac{\beta+1}{\beta}\int{p_\text{true}^t(y)g(y|x_t) ^{\beta}\mathrm{d}y}+\int{g(y|x_t)^{\beta+1}\mathrm{d}y}.
\end{aligned}    
\end{equation}
Here, $\beta \in (0, 1)$, and
\begin{equation}\nonumber
\lim_{\beta\to0}\mathcal{D}_{\beta}(\cdot||\cdot)=\mathcal{D}_{\text{KL}}(\cdot||\cdot).    
\end{equation} 
The term $\frac{1}{\beta}\int{p_\text{true}^t(y)^{\beta+1}\mathrm{d}y}$ in \eqref{eq.beta divergence} can be neglected because it is a constant value independent of $x_t$. Thus, the minimizer of $\mathcal{D}_{\beta}\left(p_\text{true}^t(\cdot)||g(\cdot|x_t)\right)$  equals
\begin{equation}\label{eq.modified beta divergence}
\begin{aligned}
\arg\min_{x_t}\Big\{ -\frac{\beta+1}{\beta}\int{p_\text{true}^t(y)g(y|x_t) ^{\beta}\mathrm{d}y}\\+\int{g(y|x_t)^{\beta+1}\mathrm{d}y}\Big\}.   
\end{aligned}
\end{equation}
% To apply the $\beta$-divergence in practice, we can't directly access the true data distribution $p_\text{true}^t(y)$ in \eqref{eq.modified beta divergence}. 
% Instead, we use the empirical data distribution $p_\text{em}^t(y)=\delta(y-y_t)$ as a replacement \cite{akrami2022robust, boustati2020generalised, futami2018variational, ghosh2016robust}. 
When applying the $\beta$-divergence in practice, since we cannot directly access the true data distribution $p_\text{true}^t(y)$ in \eqref{eq.modified beta divergence}, we use the empirical data distribution $p_\text{em}^t(y)=\delta(y-y_t)$ as replacement \cite{akrami2022robust, boustati2020generalised, futami2018variational, ghosh2016robust}. 
As a result, we can obtain the loss function $\ell^\beta(x_t,y_t)$ related to the $\beta$-divergence by substituting $p_\text{true}^t(y)$ with $p_\text{em}^t(y)$:
\begin{equation}\label{eq.beta loss}
\begin{aligned}
\ell^\beta(x_t,y_t)=-\frac{\beta+1}{\beta}g(y_t|x_t) ^{\beta}
+\int{g(y|x_t)^{\beta+1}\mathrm{d}y}.
\end{aligned}
\end{equation}
When choosing $G(y_t|x_t)$ as the ``$\beta$ likelihood" 
\begin{equation}\label{eq.general beta likelihood}
G^{\beta}(y_t|x_t):=\exp{\left(-\ell^{\beta}(x_t,y_t)\right)},
\end{equation}
we can  modify \eqref{eq.mhe-3} in the MHE objective to
\begin{equation}\label{eq.beta-mhe-3}
h(y_k,x_k)=\ell^\beta(x_k,y_k),
\end{equation} 
leading to the $\beta$-MHE algorithm. Note that the other components of the MHE algorithm are retained. Thus, the computational complexity of $\beta$-MHE aligns with that of MHE, since the only alteration lies in the \eqref{eq.mhe-3} and \eqref{eq.beta-mhe-3}.
% Furthermore, as detailed in Section \ref{sec.III}, \eqref{eq.beta-mhe-3} can be expressed as $h(\zeta_k)=\ell^\beta(\zeta_k)$ using SSM formulation of MHE.

As summarized in Algorithm \ref{alg:algorithm1}, the $\beta$-MHE algorithm achieves robustness to outliers by employing a cost function that leverages $\beta$-divergence. The choice of the parameter $\beta$ balances the robustness and accuracy of $\beta$-MHE, with smaller $\beta$ appropriate for minor measurement outliers, and larger $\beta$ suitable for significant measurement outliers. Notably, as $\beta$ approaches zero, $\beta$-MHE converges to the standard MHE, facilitating a smooth transition between the two methods.

% \begin{remark}
% Minimizing the $\beta$-divergence from the empirical distribution $p_\text{em}$ to the likelihood $g$ gives the well-known M-estimator (see \cite{futami2018variational} for details) \chang{can you briefly describe M-estimator here?}.
% \end{remark}
\begin{algorithm}
\caption{$\beta$-MHE Algorithm}\label{alg:algorithm1}
\begin{algorithmic}[1]
\Require The hidden Markov model \{$\pi_0$, $f$, $g$\}, the estimation horizon $T$, the previous state estimate $\hat{x}_{t-T|t-T}$ at time $t-T$, and the measurements $y_{t-T+1:t}$
\Ensure State estimate $\hat{x}_t$
\If{$0<t<T$}
    \State The estimation horizon is reduced to $T'=t$. Solve \eqref{eq.mhe-a},\eqref{eq.mhe-b},\eqref{eq.mhe-1},\eqref{eq.mhe-2},\textcolor{blue}{\eqref{eq.beta-mhe-3}} using the arrival cost:
    \begin{align}\nonumber
        \Gamma(x_0) = \Vert x_{0}-\hat{x}_{0|0} \Vert^2_{P^{-1}_{0}},
    \end{align}
    where $\hat{x}_0$ and $P_0$ are the mean and variance of the distribution $\pi_0$.
\Else
    \State Solve \eqref{eq.mhe-a},\eqref{eq.mhe-b},\eqref{eq.mhe-1},\eqref{eq.mhe-2},\textcolor{blue}{\eqref{eq.beta-mhe-3}} using the arrival cost:
    \begin{align}\nonumber
        \Gamma(x_{t-T}) = \Vert x_{t-T}-\hat{x}_{t-T|t-T} \Vert^2_{P^{-1}_{t-T}},
    \end{align}
    where $P_{t-T}$ is calculated via EKF or UKF.
\EndIf
\end{algorithmic}
\end{algorithm}
\section{Robustness Analysis}\label{sec.V}

% In Section \ref{sec.III}, we introduce the $\beta$-MHE algorithm, which theoretically employs the $\beta$-divergence to enhance robustness against outliers. As robustness plays a pivotal role in statistics when estimators encounter outliers \cite{hampel1986robust,hampel1974influence,rousseeuw1991tutorial,huber2011robust}, it becomes equally imperative to ensure it in the context of MHE. Consequently, verifying this theoretical claim through empirical evaluation is indispensable. For this purpose, we 
Robustness plays a pivotal role in characterizing the estimator's ability to maintain accuracy in the presence of outliers\cite{hampel1986robust,hampel1974influence,rousseeuw1991tutorial,huber2011robust}. 
To evaluate the robustness of MHE,
we utilize the influence function, a widely used measure in robust statistics determining an estimator's sensitivity to outliers \cite{hampel1986robust,hampel1974influence,huber2011robust}. 
In particular, under linear Gaussian scenarios, we evaluate the supremum of the influence function for both MHE and $\beta$-MHE. Our findings confirm that the $\beta$-MHE, exhibiting a finite influence function, indeed provides improved robustness compared to MHE, which exhibits an unbounded influence function.

% \sout{In section \ref{sec.III}, we have demonstrated that $\beta$-divergence surpasses KL divergence under measurement outliers intuitively\wei{I don't like this , nothing so intuitive.}. 
% In this section, we measure the robustness \wei{again, logic broken. Why you need to measure the robustness now?} of $\beta$-MHE via the lens of influence function\wei{an why you have to use influence function? If I understand correctly, you didn't introduce influence function in-formally?}, which quantifies the effect of infinitesimal perturbations in the data on some estimated statistics.
% Specifically, we derive the influence function of $\beta$-MHE and investigate the maximum impact of the contaminated observation for different estimators \wei{what do you mean by 'different estimators'} in the linear Gaussian case.}

\subsection{Influence Function of the MHEs}

% The Influence Function serves as a fundamental tool in robust statistics, quantifying an estimator's resilience to data contamination. Although successfully implemented in robustness analyses of the linear Kalman Filter, its application to nonlinear systems, specifically nonlinear MHE, remains a challenge due to inherent complexities. Existing MHE robustness analyses typically employ self-designed cost metrics, limiting their studies to linear systems, excluding system noise, and considering only bounded outliers. In response to this research gap, this subsection proposes the use of IF, a well-founded measure, for enhancing the robustness analysis of nonlinear MHE.
The influence function has been successfully implemented in robustness analyses of KFs \cite{gandhi2009robust,kovavcevic1992robust}. However, its application to the robustness analysis of MHE remains unexplored. This is mainly because previous works 
formulate MHE from an optimization standpoint, neglecting its essential nature as an inference problem. 
Existing robustness analyses for MHE are based on manually-designed robustness metrics \cite{alessandri2014moving,alessandri2016moving}. Unlike the influence function, these metrics lack a solid statistical foundation. Such manually-designed
metrics restricts robustness analyses to linear systems, neglects process noise, and considers only bounded outliers \cite{alessandri2014moving,alessandri2016moving}. Recognizing this research gap, this subsection proposes the adoption of the influence function as a tool to facilitate the robustness analysis of nonlinear MHE. 

As discussed in section \ref{sec.III}, through an optimization-centric view on Bayesian inference, the loss function $\ell(x,y)$ in \eqref{eq.gbi} is designed by minimizing the divergence between the empirical data distribution $p_\text{em}(y')=\delta(y'-y)$ and the likelihood function $g(y'|x)$. Suppose the empirical data distribution is contaminated by a data point $z$ with probability $\epsilon$, such that $p_\text{em}(y')=(1-\epsilon)\delta(y'-y)+\epsilon\delta(y'-z)$, the resulting estimator is then denoted as $\hat{x}\left((1-\epsilon)g+\epsilon \delta(z)\right)$ or simply as $\hat{x}^{z, \epsilon}$. Influence function is defined as the value of the derivative of $\hat{x}\left((1-\epsilon)g+\epsilon \delta(z)\right)$ with respect to the contamination probability $\epsilon$ at $\epsilon=0$:
\begin{equation}\nonumber
\left.\mathbb{IF}(z,\hat{x},g)=\frac{\partial{}}{\partial{\epsilon}}\left[\hat{x}\left((1-\epsilon)g+\epsilon \delta(z)\right)\right]\right|_{\epsilon=0}.
\end{equation}
% where $\hat{x}\left((1-\epsilon)g+\epsilon \delta(z)\right)$ represents the state estimate derived from the likelihood $g(y|x)$ with the contaminated empirical data distribution 
% $p_\text{em}(y')=(1-\epsilon)\delta(y'-y)+\epsilon\delta(y'-z)$. 
% \chang{how to understand this $p_{em}$? Does it replace $g(y|x)$?}
% While the general IF analysis has been extensively conducted in statistics \cite{huber2011robust,hampel1986robust,hampel1974influence}, there is a distinct gap in the study of the influence function of MHE. 
% This is primarily because previous works perceive MHE as an optimization problem and do not consider it from a Bayesian perspective \cite{alessandri2010advances,robertson1996moving}. The latter viewpoint is essential for a more holistic understanding of MHE's robustness, as it incorporates the influence of outliers within the data distribution. 
% \wei{why? is that because it is toooo trivial? if not, why challenging}.
Before we present our main result, we make the following assumption:
% \begin{assumption}\label{assump.continuity}
% $F(\cdot)$, $h(\cdot)$, $f(\cdot)$
% and $G(\cdot)$ are continuous.
% % The solution to \eqref{eq.mhe-a}\eqref{eq.mhe-b} exists for all $T\in\mathbb{N}$. The sufficient
% % conditions for the existence of solutions are well studied in
% % \cite{rao2003constrained}. For example, it requires the stage function $\mathbb{C}(x_t, x_{t-1}, y_t) = k(x_{t},x_{t-1})+h_t(y_t,x_t)$ to be bounded by two $K_+$-functions. \textcolor{red}{To be modified...}
% \end{assumption}
% \begin{assumption}\label{assump.stage cost}
% The stage costs $\mathbb{C}(\xi_t, \zeta_t
% )$ are continuous functions and satisfy the following inequalities,
% \begin{equation}\nonumber
% \underline{\alpha}_{\xi}(\Vert \xi_t \Vert) + \underline{\alpha}_{\zeta}(\Vert \zeta_t \Vert) \leq \mathbb{C}(\xi_t, \zeta_t
% ) \leq \bar{\alpha}_{\xi}(\Vert \xi_t \Vert) + \bar{\alpha}_{\zeta}(\Vert \zeta_t \Vert),
% \end{equation}
% in which $\underline{\alpha}_{\xi}$, $\underline{\alpha}_{\zeta}$, $\bar{\alpha}_{\zeta}$ and $\bar{\alpha}_{\zeta}$ are $\mathcal{K}_{\infty}$ functions.
% \end{assumption}
\begin{assumption}\label{assump.diffenertiable}
% The loss function $\ell(x_t,y_t)$ is twice differentiable with respect to its second argument.
$J(x_{t-T:t})$ is twice continuously differentiable with respect to $x_{t-T:t}$ \footnote{Here we suppose this assumption holds for both the uncontaminated and contaminated empirical distributions.} and its second derivative at the local minimum is nonsingular.
\end{assumption}
% \wei{use Remark on these assumptions}
\begin{remark}
Assumption \ref{assump.diffenertiable} allows the derivation of an analytical form of the influence function. Assumption \ref{assump.diffenertiable} is not overly restrictive, as it only necessitates a nonsingular Hessian matrix at the local minimum point. Besides, the twice continuous differentiability condition is naturally fulfilled in the case of MHE and $\beta$-MHE for linear Gaussian systems, which can be validated through the formulation of stage costs in \eqref{eq.mhe for linear Gaussian systems} and \eqref{eq.h(zeta) for beta MHE}.
\end{remark}
% \wei{so, only use assumption 3, find analytically is trivial...}.
The analytical expression of the influence function is presented in Theorem \ref{theorem.IF of MHE}.
\begin{theorem}\label{theorem.IF of MHE}
Under Assumption \ref{assump.well-posed} and \ref{assump.diffenertiable}, the influence function of $\beta$-MHE for a general system \eqref{eq.hmm} can be written as
\begin{subequations}
\begin{align}
&\mathbb{IF}(z, \hat{x}_{t-T: t}, G^{\beta}(y_t|x_t))=M_1^{-1}(\hat{x}_{t-T: t})\cdot M_2(z,\hat{x}_{t-T: t}), \label{eq. IF of beta MHE}
\\
&M_1(\hat{x}_{t-T: t}) := \left.\frac{\partial^2{\left[J(x_{t-T:t})\right]}}{\partial{x_{t-T:t}^2}}\right|_{x_{t-T:t}=\hat{x}_{t-T:t}}, \label{eq. M1 matrix}
\\
&M_2(z, \hat{x}_{t-T: t}) := \sum\limits_{i=t-T+1}^t\left.\frac{\partial{\left[\mathscr{K}^{\beta, z}(x_i,y_i)\right]}}{\partial{x_{t-T:t}}}\right|_{x_{t-T:t}=\hat{x}_{t-T:t}}, \label{eq. M2 matrix}
\end{align}
\end{subequations}
where $J(x_{t-T:t})$ is defined by \eqref{eq.mhe-b}\eqref{eq.mhe-1}\eqref{eq.mhe-2}\eqref{eq.beta-mhe-3} and $\mathscr{K}^{\beta, z}(x_t,y_t)=\frac{\beta+1}{\beta}{g(y_t|x_t)^{\beta}}-\frac{\beta+1}{\beta}{g(z|x_t)}^{\beta}$. 
\end{theorem}
The proof of Theorem \ref{theorem.IF of MHE} can be found in Appendix \ref{appendix.theorem 1}.
\begin{corollary}
Similar results hold for the MHE when $\mathscr{K}^{\beta,z}(x_t,y_t)$ is restricted to $\mathscr{K}^z(x_t,y_t)=\log{g(y_t|x_t)}-\log{g(z|x_t)}$ and  $J(x_{t-T:t})$ is defined by \eqref{eq.mhe-b}\eqref{eq.mhe-1}\eqref{eq.mhe-2}\eqref{eq.mhe-3}.
\end{corollary}
\begin{proof}
Let $\beta \to 0$, then the $\beta$-MHE converges to MHE, and therefore the results can be obtained for MHE.
\end{proof}

\subsection{Gross Error Sensitivity
of MHEs for Linear Guassian Systems}
% \wei{logic, why you now discuss sensitivity now}
% \han{In the previous subsection, we derived the influence function of $\beta$-MHE and MHE. Notably, although the influence function evaluates the robustness of an estimator under a specific point of contamination, our goal is to assess the estimator's robustness comprehensively, without reliance on a specific contamination point. To this end, we focus on worst-case scenarios, namely considering the supremum of the influence function across all possible contamination points, as the boundedness of this supremum would imply the estimator's robustness. This introduces us to the concept of gross error sensitivity.}
The previous subsection derives the influence function of $\beta$-MHE and MHE under a specific point of contamination. To comprehensively assess the estimator's robustness, this subsection focuses on worst-case scenarios, namely considering the supremum of the influence function across all possible contamination points. This leads to the concept of gross error sensitivity \cite{ko1988robustness}.

Gross error sensitivity measures the maximum change a small perturbation to the likelihood $g$ at a point $z$ can induce to the estimate $\hat{x}$. It is defined as 
\begin{equation}\nonumber
{\gamma}(\hat{x},g) = \sup_{z}\Vert \mathbb{IF}(z,\hat{x},g) \Vert.    
\end{equation}
We consider the typical linear Gaussian systems
\begin{equation}\label{eq.linear Guassian model}
\begin{aligned}
& x_{t+1} = Ax_t + \xi_t, \quad y_t = Cx_t + \zeta_t,
\\
& \xi_t \sim \mathcal{N}(0,\;Q), \quad \zeta_t \sim \mathcal{N}(0,\;R), \quad
x_0 \sim \mathcal{N}(0,\;P_0),
\end{aligned}
\end{equation}
where $x_t \in \mathbb{R}^n$, $y_t \in \mathbb{R}^m$, $Q \succ 0$ and $R \succ 0$. The terms $\xi_t$ and $\zeta_t$ refer to the additive process noise and measurement noise, respectively. The arrival cost and the stage cost functions of MHE for system \eqref{eq.linear Guassian model} are defined as
\begin{equation}\label{eq.mhe for linear Gaussian systems}
\begin{aligned}
\Gamma(x_{t-T})&=\frac{1}{2}\Vert x_{t-T}-\hat{x}_{t-T|t-T} \Vert^2_{P_{t-T}^{-1}},\\
k(x_t,x_{t-1})&=\frac{1}{2}\Vert x_{t}-Ax_{t-1} \Vert^2_{{Q}^{-1}},\\
h(y_t,x_t)&=\frac{1}{2}\Vert y_t-Cx_t \Vert^2_{{R}^{-1}}.
\end{aligned}
\end{equation}
Here, $\hat{x}_{t-T|t-T}$ represents the optimal estimate from the previous time and $P_{t-T}$  refers to the error covariance matrix, which can be calculated recursively by solving the matrix Riccati equation \cite{rao2003constrained}
\begin{equation}\label{eq.riccati equation}\nonumber
\begin{aligned}
P_{i+1} = Q + AP_iA^{\top} - AP_iC^{\top}(R+CP_iC^{\top})^{-1}CP_iA^{\top}.
\end{aligned}    
\end{equation}
The problem setting for the $\beta$-MHE method is largely similar to that of  MHE,  with the only adjustment being the modification of $h(y_t,x_t)$ in \eqref{eq.mhe for linear Gaussian systems}. Specifically, by substituting \eqref{eq.mhe-3} with \eqref{eq.beta-mhe-3}, $h(y_t,x_t)$ in \eqref{eq.mhe for linear Gaussian systems} is altered to
\begin{equation}\label{eq.h(zeta) for beta MHE}
\begin{aligned}
h(y_t,x_t)&=-\frac{\beta+1}{\beta \sqrt{(2\pi)^{\beta m}|R|^{\beta}}}e^{-\frac{\beta}{2}\Vert y_t-Cx_t \Vert^2_{{R}^{-1}}}\\
&+\frac{1}{(\beta+1)^{\frac{1}{2}}(2\pi)^{\frac{m\beta}{2}}|R|^{\frac{\beta}{2}}}.
\end{aligned}
\end{equation}

The subsequent theorem characterizes the robustness properties of MHE methods for linear Gaussian systems.
\begin{theorem}\label{theorem.bounded IF}
Under Assumption \ref{assump.well-posed} and \ref{assump.diffenertiable}, the gross error sensitivity of MHE for system \eqref{eq.linear Guassian model} is infinite while for that of $\beta$-MHE, it is bounded. Specifically, we have
\begin{equation}\label{eq.bound for beta MHE}
\begin{aligned}
\gamma(\hat{x}_{t-T: t}, G^{\beta}(y_t|x_t))&\leq 2\sqrt{T} \cdot\Vert M_1^{-1}(\hat{x}_{t-T: t}) \Vert_{\mathcal{F}}\cdot \rho_{\text{max}} ,
\end{aligned}
\end{equation}
where $M_1(\hat{x}_{t-T: t})$ is defined as \eqref{eq. M1 matrix}, and 
\begin{equation}\label{eq.definition of rho}
\begin{aligned}
\rho_{\text{max}} &:= \max_{i \in \mathbb{Z}_{[t-T, t]}} \sup_z \Vert \rho(\hat{x}_i,z) \Vert,\\
\rho(\hat{x}_t,z) &:=   \frac{(\beta+1)C^{\top}R^{-1}}{\sqrt{(2\pi)^{\beta m}|R|^{\beta}}}(z-C\hat{x}_t)e^{-\frac{\beta}{2}\Vert z-C\hat{x}_t \Vert^2_{{R}^{-1}}}.
\end{aligned}
\end{equation}
\end{theorem}
The proof of Theorem \ref{theorem.bounded IF} can be found in Appendix \ref{appendix.theorem 2}.
\begin{remark}
Due to the selected arrival cost, choosing the MHE horizon length $T = 1$ recovers
KF. Therefore, Theorem \ref{theorem.bounded IF} also suits the robustness analysis of KF.
\end{remark}
Theorem \ref{theorem.bounded IF} provides an important conclusion: for linear Gaussian systems, the supremum of the influence function is finite for $\beta$-MHE, marking a distinct contrast to MHE, for which the supremum is infinite. This underscores the superior robustness offered by $\beta$-MHE over MHE.
% \wei{it is nice to give a brief summary here}

\section{Stability Analysis}\label{sec.VI}
% \wei{logic, why stability??}
The stability of MHE plays a significant role in its theoretical foundation, as it guarantees the estimates' convergence to the true state \cite{rao2003constrained, hu2021generic, schiller2022lyapunov,allan2021robust,knufer2023nonlinear,battistelli2017moving}. This section analyzes the stability of $\beta$-MHE inspired by the result in \cite{allan2019moving}. 
% In this section, we embark on a stability analysis of $\beta$-MHE. 
As a preliminary step, we introduce the concept of incremental input/output-to-state stability (i-IOSS) based on \eqref{eq.IOSS},  which will play a key role in our subsequent analysis.
% \wei{this section is a good theoretical contribution, you should highlight that! Though techniques-wise, not surprisingly talented}
% \begin{definition}[i-IOSS% and exp i-IOSS
% ]\label{def.ioss}
% The system $x_{t+1}=F(x_t, \xi_t)$, $y_t=H(x_t)$ is i-IOSS if there exist the $\mathcal{KL}$ function $\alpha_0$ and $\mathcal{K}$ function $\alpha_{\xi}$  and $\alpha_{\zeta}$ which satisfy
% \begin{equation}\label{eq.IOSS}
% \begin{aligned}
% \Vert x_{t}^{(1)}-x_{t}^{(2)}\Vert &\leq \alpha\left(\Vert x_0^{(1)}-x_0^{(2)}\Vert, t\right)\\&+\alpha_{\xi}\left(\max_{0 \leq i \leq t-1}\Vert \xi_i^{(1)}-\xi_i^{(2)}\Vert\right)\\&+\alpha_{\zeta}\left(\max_{0 \leq i \leq t-1}\Vert H(x_i^{(1)})-H(x_i^{(2)})\Vert\right)
% ,\;\forall t>0.
% \end{aligned}
% \end{equation}
% \end{definition}
\begin{definition}[i-IOSS% and exp i-IOSS
]\label{def.i-ioss}
The system $x_{t+1}=\mathcal{F}(x_t, \xi_t)$, $y_t=\mathcal{G}(x_t) + \zeta_t$ is i-IOSS if there exist $\mathcal{KL}$ function $\gamma_x(\cdot,\cdot)$, and $\mathcal{K}$ functions
$\gamma_{\xi}(\cdot), \gamma_{\zeta}(\cdot)$ which satisfy
\begin{equation}
\begin{aligned}
\left\| x_{t}^{(1)}-x_{t}^{(2)} \right\|
&\leq \gamma_x\left( \left\| x_0^{(1)}-x_0^{(2)} \right\|, t\right)  \\
&\oplus  \gamma_{\xi} \left(\left\| \xi^{(1)} - \xi^{(2)}
\right\|_{0:t-1} \right)
\\
&\oplus  \gamma_{\zeta} \left(\left\| y^{(1)}-y^{(2)} 
\right\|_{1:t} \right)
,\;\forall t\in\mathbb{Z}_{[0,\infty)}.
\label{eq.IOSS}
\end{aligned}
\end{equation}
Here, $x_t^{(1)}$ and $x_t^{(2)}$ are two state sequences with initial states $x_0^{(1)}$ and $x_0^{(2)}$ respectively.
\end{definition}
The concept of i-IOSS allows us to measure a system's resistance to external perturbations.
As summarised in Assumption \ref{assump.contraction map}, we assume the underlying system to be IOSS and the $\mathcal{KL}$ function $\gamma_x$ eventually become a contraction map on all intervals from the origin to $\bar{s}$.

\begin{assumption} \label{assump.contraction map}
The system $x_{t+1}=\mathcal{F}(x_t, \xi_t)$, $y_t=\mathcal{G}(x_t) + \zeta_t$ is i-IOSS and furthermore, for every $\Bar{s} > 0$ and for every $\eta \in (0, 1)$, there exists some $T_m \geq 0$ such that
\begin{equation}
\begin{aligned}
\gamma_x(s, T_m) \leq \eta s,    
\end{aligned}
\end{equation}
$\forall$ $s \leq \Bar{s}$.   
\end{assumption}
To further study the system's stability properties, we proceed to introduce the definition of robustly asymptotically stable (RAS).
%robust asymptotic stability (RAS).
\begin{definition}[RAS \cite{allan2019moving}]\label{def.robust state estimation}
An estimator is RAS if there exist $\bar{s}, \delta> 0$,
$\mathcal{KL}$ functions $\rho_x(\cdot,\cdot)$, and $\mathcal{K}$ functions
$\rho_{\xi}(\cdot), \rho_{\zeta}(\cdot)$
such that if $\|x_0 - \hat{x}_{0|0}\| \leq \Bar{s}$, $\| \xi \|_{0: \infty}, \| \zeta \|_{0: \infty} \leq \delta$, then for $\forall t\in \mathbb{Z}_{[0,\infty)}$, we have
\begin{equation}\nonumber
\begin{aligned}
\left\| x_{t} - \hat x_{t|t} \right\| &\le \rho_x \left( \left \| x_0 - \hat{x}_{0|0} \right \| ,t \right) 
\\
&\oplus \rho_\xi \left(
\left\| \xi \right\|_{0:t-1} \right) 
\\
&\oplus \rho_\zeta \left(
\left\| \zeta \right\|_{1:t}\right).
\end{aligned}    
\end{equation}
\end{definition}

To establish the RAS of MHE, we lay down two crucial assumptions.
\begin{assumption}\label{assump.bounded cost function}
The stage costs $k(\xi_t
)$ and $h(\zeta_t)$ are bounded by $\mathcal{K}$ functions, i.e.,
\begin{equation}\label{eq.bounded stage costs}
\begin{aligned}
\underline{\gamma}_{\xi}{(\|\xi_t\|)} & \leq k(\xi_t) \leq \Bar{\gamma}_{\xi}{(\|\xi_t\|)}, \\ 
\underline{\gamma}_{\zeta}{(\|\zeta_t\|)} & \leq h(\zeta_t) \leq \Bar{\gamma}_{\zeta}{(\|\zeta_t\|)}.    
\end{aligned}
\end{equation}
Here, $\underline{\gamma}_{\xi}, \Bar{\gamma}_{\xi}, \underline{\gamma}_{\zeta}, \Bar{\gamma}_{\zeta}$ are $\mathcal{K}$ functions.
\end{assumption}
\begin{assumption} \label{assump.contraction map2}
For every $\Bar{s} > 0$ and $\eta \in (0, 1)$, there exists $\mu > 0$ such that
$\gamma_{\xi} \left( 2\underline{\gamma}_{\xi}^{-1}
\Big(
3\mu \cdot s^2 
\Big) \right) \leq \eta s$,
$\gamma_{\zeta}\left( 2\underline{\gamma}_{\zeta}^{-1}
\Big(
3\mu \cdot s^2 
\Big) \right)  \leq \eta s$, 
$\forall$ $s \leq \Bar{s}$.   
\end{assumption}
\begin{remark}
Assumption \ref{assump.bounded cost function} assumes that the cost functions are bounded by $\mathcal{K}$ functions. This is a natural assumption that can be easily verified in most scenarios. For example, for the system with Gaussian noises, the stage costs $k(\xi_t) = \frac{1}{2} \| \xi_t \|^2_{Q^{-1}}$ and $h(\zeta_t)=\ell^{\beta}(\zeta_t) = -\frac{\beta+1}{\beta \sqrt{(2\pi)^{\beta m}|R|^{\beta}}}e^{-\frac{\beta}{2}\Vert \zeta_t \Vert^2_{{R}^{-1}}}$ can be verified to satisfy \eqref{eq.bounded stage costs} by noticing:
\begin{equation}\nonumber
\begin{aligned}
\frac{1}{2}\underline{\lambda}_Q \| \xi_t \|^2 &\leq \frac{1}{2} \| \xi_t \|^2_{Q^{-1}} \leq \frac{1}{2}\bar{\lambda}_Q \| \xi_t \|^2, 
\\
-e^{-\frac{\beta}{2} \underline{\lambda}_{R} \Vert \zeta_t \Vert^2}
&\leq
-e^{-\frac{\beta}{2}\Vert \zeta_t \Vert^2_{R^{-1}}}
\leq
-e^{-\frac{\beta}{2} \Bar{\lambda}_R\Vert \zeta_t \Vert^2},
\end{aligned}
\end{equation}
where $\underline{\lambda}_Q, \bar{\lambda}_Q > 0$ are the minimum and maximum eigenvalues of the matrix $Q^{-1}$, and $\underline{\lambda}_R, \bar{\lambda}_R > 0$ are the minimum and maximum eigenvalues of the matrix $R^{-1}$.

\end{remark}

Following these assumptions, the stability of the corresponding $\beta$-MHE can be established:
\begin{theorem}\label{theorem.stability}
Under Assumption \ref{assump.contraction map}, \ref{assump.bounded cost function} and \ref{assump.contraction map2}, for every $\Bar{s} > 0$, there exists a $T_m$ and $\Bar{s}$ such that if $T \geq T_m$ and $\| \hat{x}_{0|0} - x_0 \| \leq \Bar{s}$, then there exist $\mathcal{KL}$ function $\rho_x$, $\mathcal{K}$ function $\rho_{\xi}$ and $\rho_{\zeta}$, and $\delta>0$, such that if $ \| \xi \|_{0:\infty}, \| \zeta \|_{0:\infty} \leq \delta$, then
\begin{equation}
\|x_t - \hat{x}_{t|t} \| \leq \rho_{x}(\|x_0 - \hat{x}_{0|0} \|, t) \oplus \rho_{\xi}(\| \xi \|_{0:t-1})  \oplus \rho_{\zeta}(\| \zeta \|_{1:t})     
\end{equation}
for $\forall t \in \mathbb{Z}_{[0, \infty)}$.    
\end{theorem}
Theorem \ref{theorem.stability} builds RAS of $\beta$-MHE and its proof can be found in Appendix \ref{appendix.Theorem 3}.
 
In this section, we establish a sufficient condition for the stability of $\beta$-MHE. We find that to ensure stability in $\beta$-MHE, the system needs to exhibit IOSS properties, the cost function should be bounded by $\mathcal{K}$ functions with contraction properties, and the estimation length has to be sufficiently large.  The stability analysis guarantees the convergence of the estimation error when performing the $\beta$-MHE algorithm.

% \wei{I expect a remark here to summarize and give some comments here, to emphasize your contributions, again and again}
\section{Simulations}\label{sec.VII}
% \wei{what questions to answer and what things do you want to prove?}

In this section, our aim is to assess the robustness of the $\beta$-MHE algorithm in the presence of outliers, as well as to investigate the effect of the $\beta$ parameter on this robustness. To this effect, we carry out two numerical simulations that showcase the performance of the algorithm. Throughout these simulations, we use the root mean squared error (RMSE) as an indicator of estimation accuracy. The RMSE measures the differences between the state estimation $\hat{x}_t$ and the actual state $x_t$:

\begin{equation}\nonumber
\begin{aligned}
\text{RMSE} = \sqrt{\frac{\sum_{t=1}^{N_\text{step}}\Vert x_t - \hat{x}_t \Vert^2}{n \cdot N_{\text{step}}}}.
\end{aligned}
\end{equation}

In this equation, $N_\text{step}$ refers to the duration of the trajectory, and $n$ is the dimension of the state. The simulations are conducted on a laptop equipped with an AMD Ryzen 7 processor and 16.0 GB RAM. The optimization problems are handled using CasADi \cite{Andersson2019}, an open-source solver for the efficient execution of nonlinear optimization tasks.
% where $N_\text{step}$ is the time length of the trajectory and $n$ is the dimension of the state. Simulations are run on a laptop equipped with AMD Ryzen 7 processor, 16.0 GB RAM, and 512 GB SSD. Meanwhile, the optimization problems are formulated
% in CasADi \cite{Andersson2019} \wei{optimization problems can be formulated in a solver.. re-write}, which is an open-source nonlinear optimization tool.
\subsection{Linear System Case: Wiener Velocity Model}
% \wei{logic, why you start with this example, motivation?} \wei{which classic filtering paper use this example? cite}
To initially test the robustness of our $\beta$-MHE algorithm in a linear system, we opt to use the Wiener velocity model, a commonly referenced model in filtering literatures \cite{sarkka2006recursive,tronarp2019student,huang2020novel}. Adopting a discretization step of $\Delta t = 0.1$, we obtain a linear Gaussian stochastic model represented by \eqref{eq.linear Guassian model}. The state is expressed as $x=\left[\begin{matrix}p_x &p_y&\Dot{p}_x&\Dot{p}_y\end{matrix}\right]^\top$, encompassing the position coordinates $p_x$, $p_y$ and the velocities $\Dot{p}_x$, $\Dot{p}_y$.
The parameters defined in \eqref{eq.linear Guassian model} are shown as
\begin{equation}\nonumber
\begin{aligned}
A &= \begin{bmatrix}
1&0&\Delta t &0\\
0&1&0&\Delta t\\
0&0&1&0\\
0&0&0&1
\end{bmatrix},\;
C = \begin{bmatrix}
1&0&0&0\\
0&1&0&0
\end{bmatrix},
\end{aligned}
\end{equation}
\begin{equation}\nonumber
\begin{aligned}
Q&=\begin{bmatrix}
\frac{\Delta t^3}{3}&0&\frac{\Delta t^2}{2} &0\\
0&\frac{\Delta t^3}{3}&0&\frac{\Delta t^2}{2}\\
\frac{\Delta t^2}{2}&0&\Delta t&0\\
0&\frac{\Delta t^2}{2}&0&\Delta t
\end{bmatrix},\;R=\mathbb{I}_{2\times 2}.
\end{aligned}
\end{equation}
To emulate the real-world scenario where measurements can be contaminated, we introduce noise following $\mathcal{N}(0, 100^2 \cdot \mathbb{I}_{2\times 2})$ with a probability $p_c$. To validate the robustness and performance of our proposed $\beta$-MHE algorithm, we compare it with MHE and KF across a varied range of $\beta$ values. The RMSE is evaluated over 100 Monte Carlo trials each consisting of 200 steps, i.e.,  $N_\text{step}=200$.

\begin{figure}[!htb]
\centering
\includegraphics[width=0.98\linewidth]{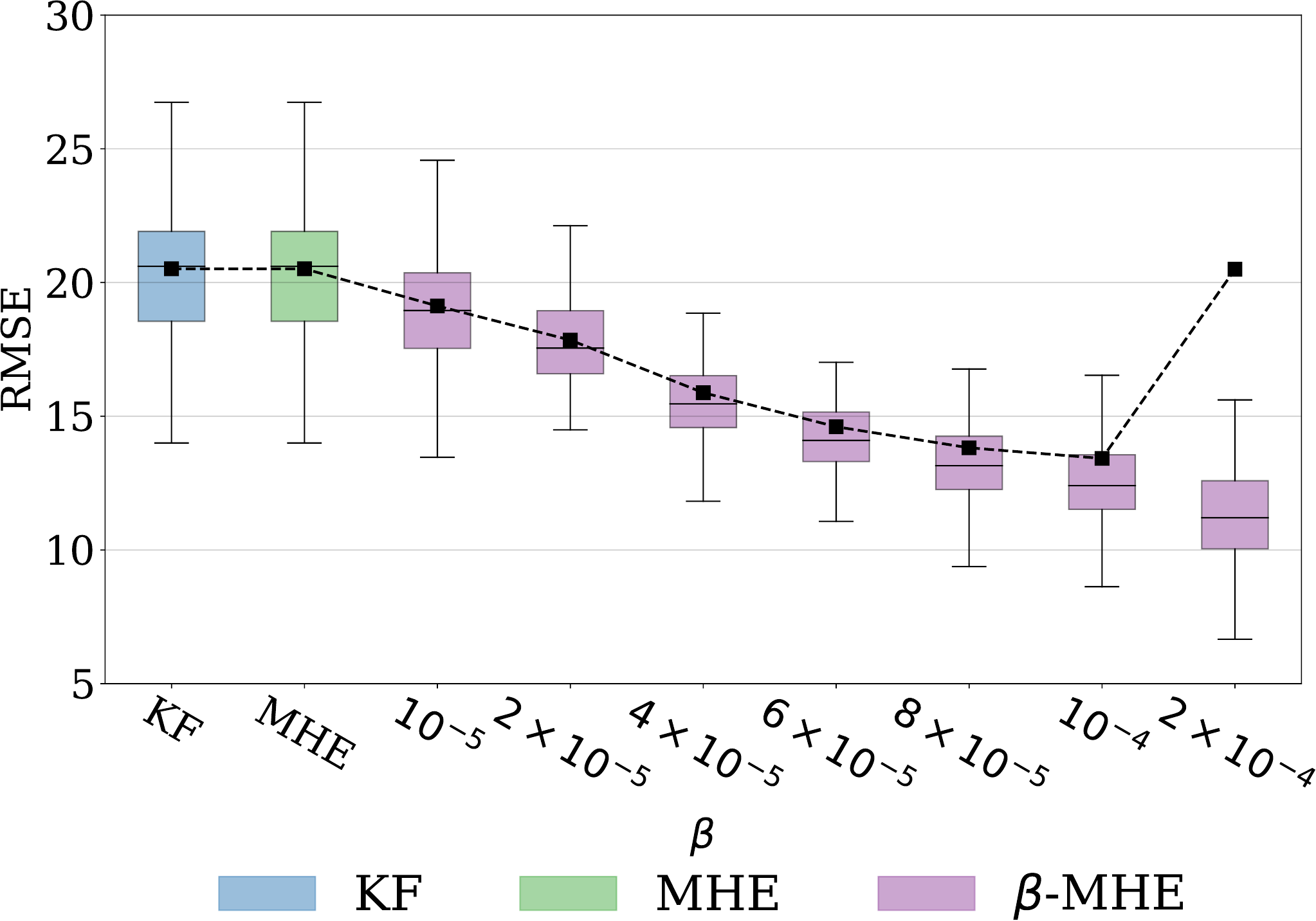}
\caption{Box plot of RMSE for KF, MHE and $\beta$-MHE with different values of $\beta$ when contamination probability $p_c=0.2$. The black square `` $\blacksquare$ " represents the average RMSE.}
\label{fig.Wiener Velocity Model}
\end{figure}

Fig. \ref{fig.Wiener Velocity Model} displays the box plot of RMSE for different methodologies at $p_c = 0.2$. Interestingly, we observe that the average RMSE demonstrates a decreasing trend as $\beta$ increases, until it attains its minimum at $\beta=10^{-4}$. This $\beta$ value emerged as a threshold from extensive experiments aiming to strike a balance between accuracy and robustness. Beyond this point, the RMSE begins to climb. Remarkably, when $\beta\leq10^{-4}$, our proposed $\beta$-MHE achieves higher estimation accuracy compared to both the KF and the conventional MHE.

\begin{figure}[!htb]
\centering 
% \captionsetup[subfigure]{justification=centering}
\subfloat[]{\includegraphics[width=0.49\linewidth]{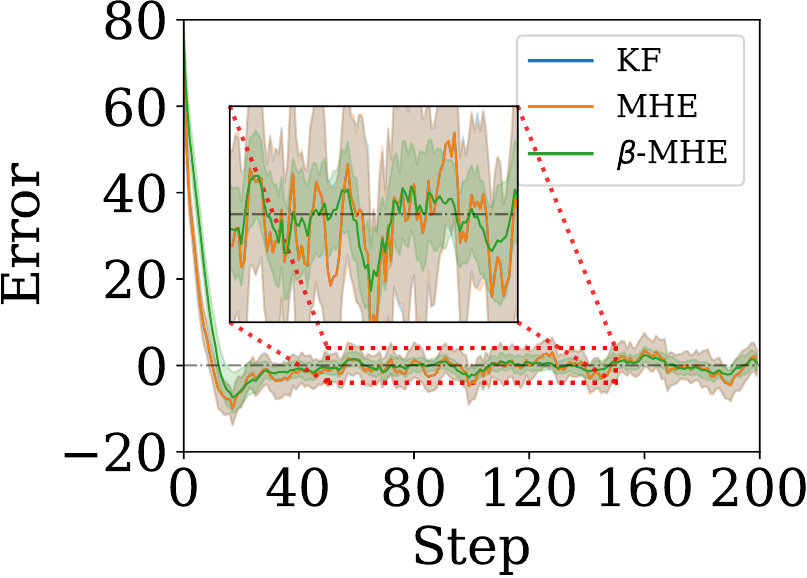}}
\subfloat[]{\includegraphics[width=0.49\linewidth]{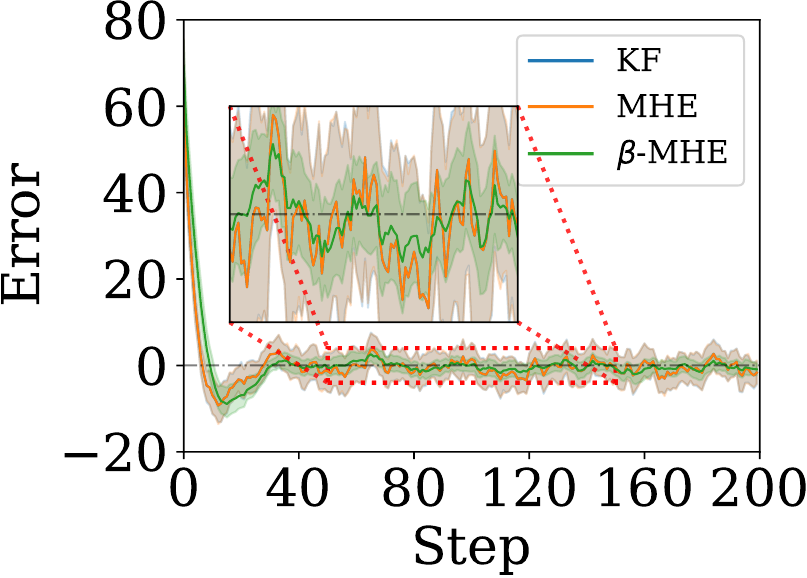}}
\\
\subfloat[]{\includegraphics[width=0.49\linewidth]{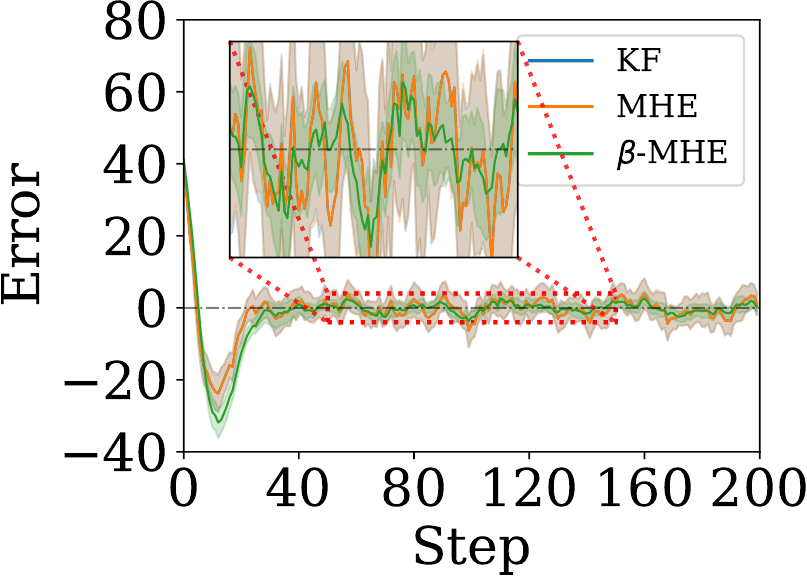}}
\subfloat[]{\includegraphics[width=0.49\linewidth]{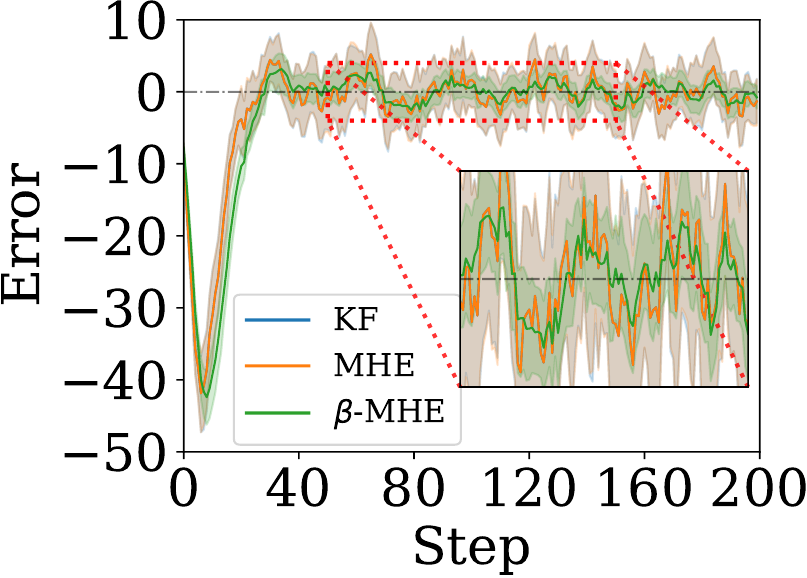}}
\caption{(a) Estimation error of
$p_x$. (b) Estimation error of
$p_y$. (c) Estimation error of
$\dot{p}_x$. (d) Estimation error of
$\dot{p}_y$. The
solid lines correspond to the mean and the shaded
regions correspond to $95\%$ confidence interval over 100
runs. Given that the MHE horizon length T is chosen as 1, the result of MHE, illustrated as the blue line, is identical to that of the KF, depicted by the orange line. }
\label{fig.problem_1 error}
\end{figure}

For further clarity, Fig. \ref{fig.problem_1 error} presents the estimation error of the four states for different methods, where $\beta$ is set as $10^{-4}$. From a computational perspective, the average processing time per step for the KF, MHE, and $\beta$-MHE were recorded as 0.00023 ms, 0.024 ms, and 0.028 ms respectively. This indicates that the computational demand of our proposed $\beta$-MHE is comparable to that of the MHE.

\subsection{Nonlinear System Case: Isothermal Gas-phase Reactor Model}
%\wei{again, why this example, which paper use this example}

\begin{figure}[!htb]
\centering
\includegraphics[width=0.98\linewidth]{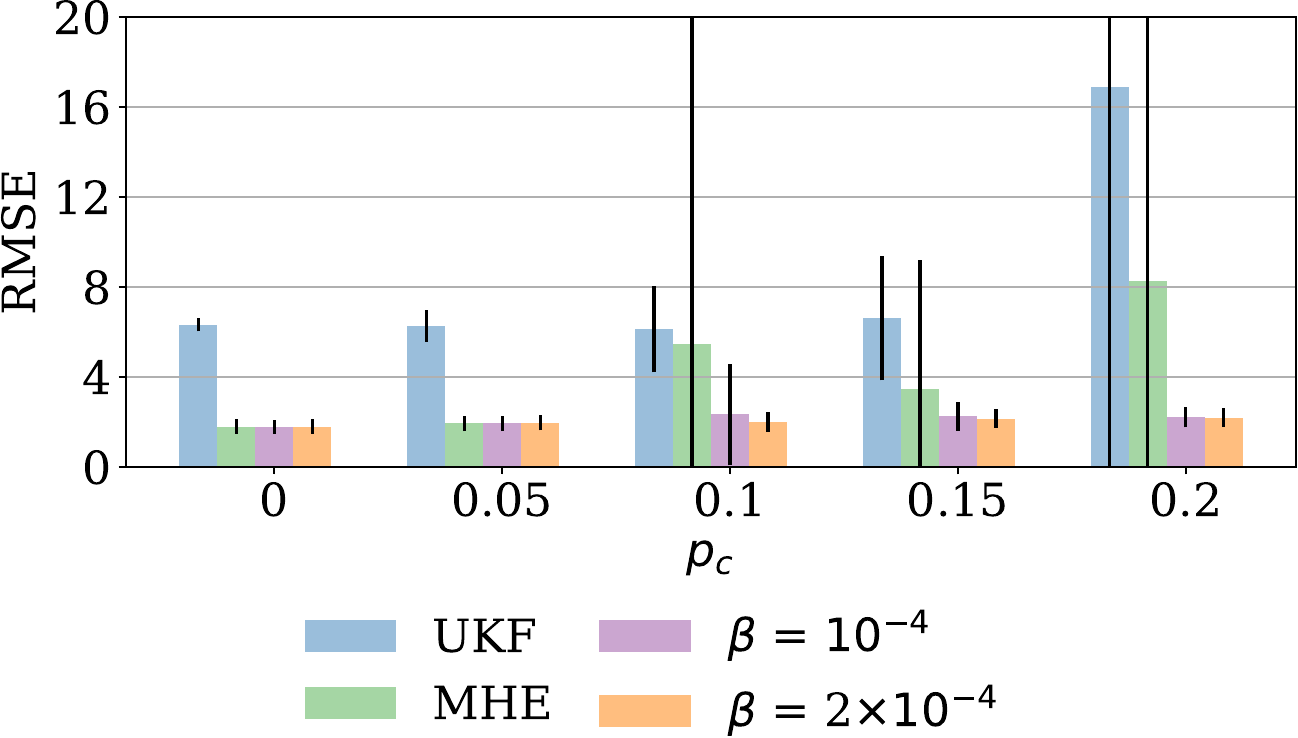}
\caption{Bar plot of RMSE for UKF, MHE, and $\beta$-MHE with different contamination probabilities. The height of each bar represents the average RMSE value for a particular method under a specific contamination level. The error bars extending from the top of each bar indicate the standard deviation of the RMSE values over 100 runs, representing the variability in performance for each method. In addition, the MHE horizon length $T$ equals 3.}
\label{fig.Reactor Model}
\end{figure}

\begin{figure}[!htb]
\centering 
% \captionsetup[subfigure]{justification=centering}
\subfloat[]{\includegraphics[width=0.49\linewidth]{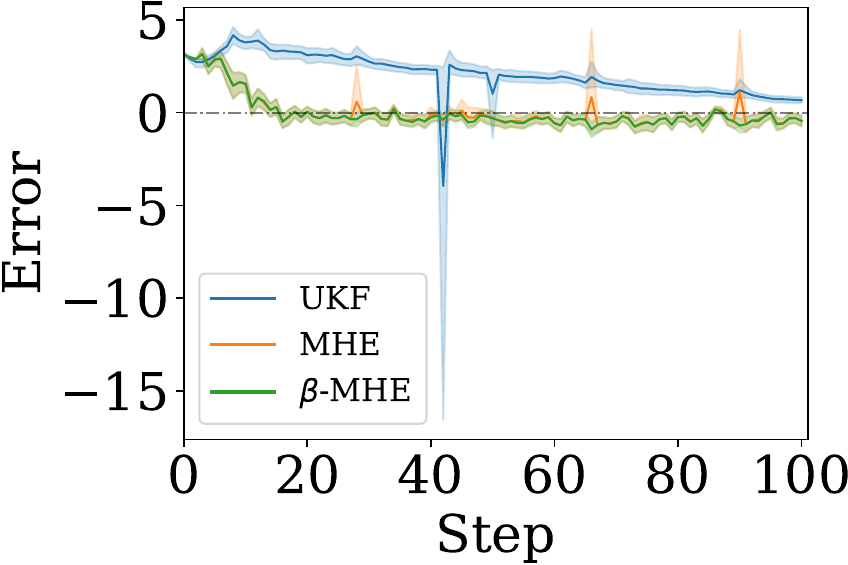}}
\subfloat[]{\includegraphics[width=0.49\linewidth]{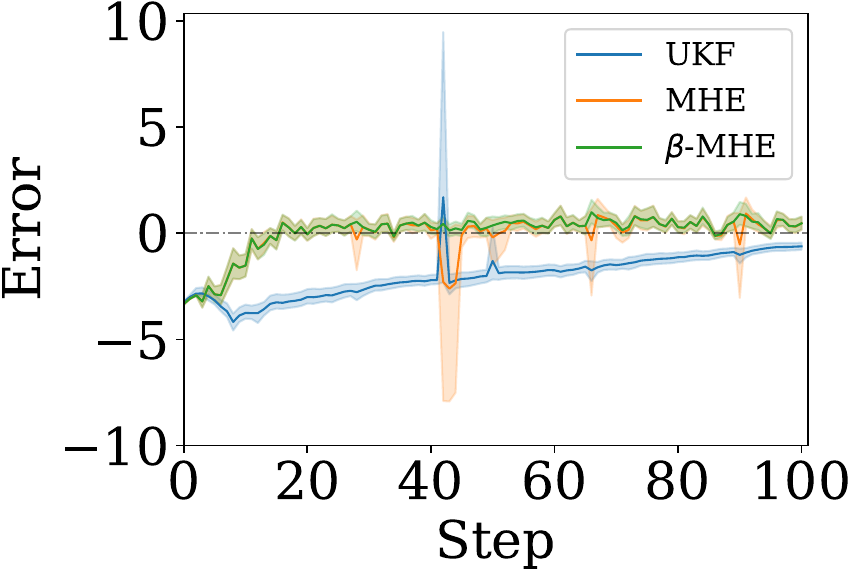}}
\caption{(a) Estimation error of
$P_A$. (b) Estimation error of
$P_B$. The
solid lines correspond to the mean, and the shaded
regions correspond to $95\%$ confidence interval over 100
runs.}
\label{fig.problem_2 error}
\end{figure}

In this subsection, we perform simulation on an isothermal gas-phase reactor model, a model commonly referenced in MHE literature \cite{1025355,gharbi2021anytime,gharbi2020proximity,schiller2022lyapunov}. This model describes the reversible reaction $2A_r \rightleftharpoons B_r$. Initially, the reactor is charged with certain amounts of $A_r$ and $B_r$, but the exact composition of the original mixture remains uncertain. The state $x$ includes the partial pressures, i.e.,   $x=\left[\begin{matrix}P_A &P_B\end{matrix}\right]^\top$. The discrete-time version of the gas-phase reactor model with the one-step Euler method  ($\Delta t = 0.1$) is
\begin{equation}\label{eq.Reactor
Model}\nonumber
\begin{aligned}
{P}_{A,t+1}&={P}_{A,t}+\left(-2k_1P_{A,t}^2+2k_2P_{B,t}\right) \cdot \Delta t + \xi_{1,t},\\
{P}_{B,t+1}&={P}_{B,t}+\left(k_1{P}_{A,t}^2-k_2{P}_{B,t}\right)\cdot \Delta t + \xi_{2,t},
\end{aligned}
\end{equation}
where $k_1 = 0.16$, $k_2 = 0.0064$, $\xi_t \sim \mathcal{N}(0,\; 10^{-4}\mathbb{I}_{2\times2})$, and $x_0 \sim \mathcal{N}(0,\; \mathbb{I}_{2\times2})$. 
A pressure gauge measures the total pressure of
the system as the species react, i.e.,
\begin{equation}\nonumber
y_t = P_{A,t} + P_{B,t} + \zeta_t,
\end{equation}
where $\zeta_t \sim \mathcal{N}(0,\; 0.01)$. It is proved in \cite{schiller2022lyapunov} that this system has IOSS property.
We simulate measurement
outliers as the impulsive noise drawing from a Student’s t distribution with
$\nu = 1$ degrees of freedom. Strictly speaking, we define $\zeta_t \sim p_c \cdot t_{\nu=1}(0,\;0.01) + (1-p_c)\cdot \mathcal{N}(0,\;0.01)$. We evaluate the results for our proposed $\beta$-MHE, the MHE, and UKF for different contamination probabilities. As shown in Fig. \ref{fig.Reactor Model}, our method surpasses UKF and the MHE under different contamination probabilities and remains effective when there are no outliers. Besides, the estimation error of the two states is illustrated in Fig. \ref{fig.problem_2 error}, where $\beta$ is chosen as $10^{-4}$ and $p_c = 0.2$. It is noticeable that the estimation error of UKF and MHE varies considerably, whereas for $\beta$-MHE, it remains relatively stable. Besides, the average
computation time of each step for UKF, the MHE, and
the $\beta$-MHE are 0.015 ms, 0.451 ms, and 0.495 ms respectively,
which reveals the computational burdens of the proposed
$\beta$-MHE and the MHE are at the same level for nonlinear systems. 


\section{Experiment: Warehouse Vehicle Localization}\label{sec.VIII}
In this section, we carry out a real-world experiment involving indoor vehicle localization to assess the robustness and accuracy of our proposed algorithm. The experiment is conducted on a  warehouse vehicle outfitted with a Lidar sensor, which supplies the necessary measurement data for real-time estimation of the vehicle's position and heading angle.

\subsection{Experimental Setup}
\begin{figure}[!htb]
\centering 
% \captionsetup[subfigure]{justification=centering}
\subfloat[]{\includegraphics[width=0.455\linewidth]{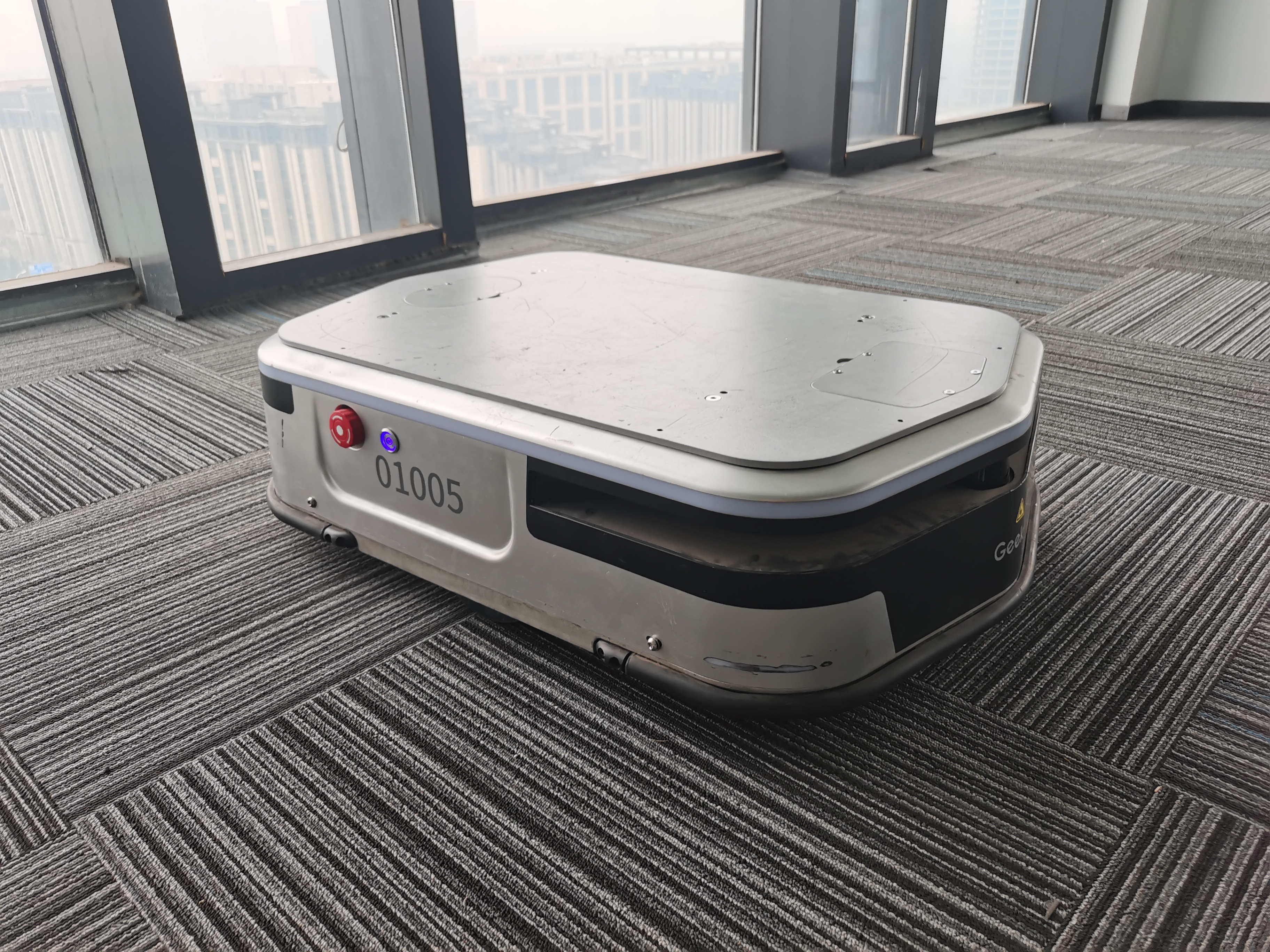}}\hspace{1mm} 
\subfloat[]{\includegraphics[width=0.490\linewidth]{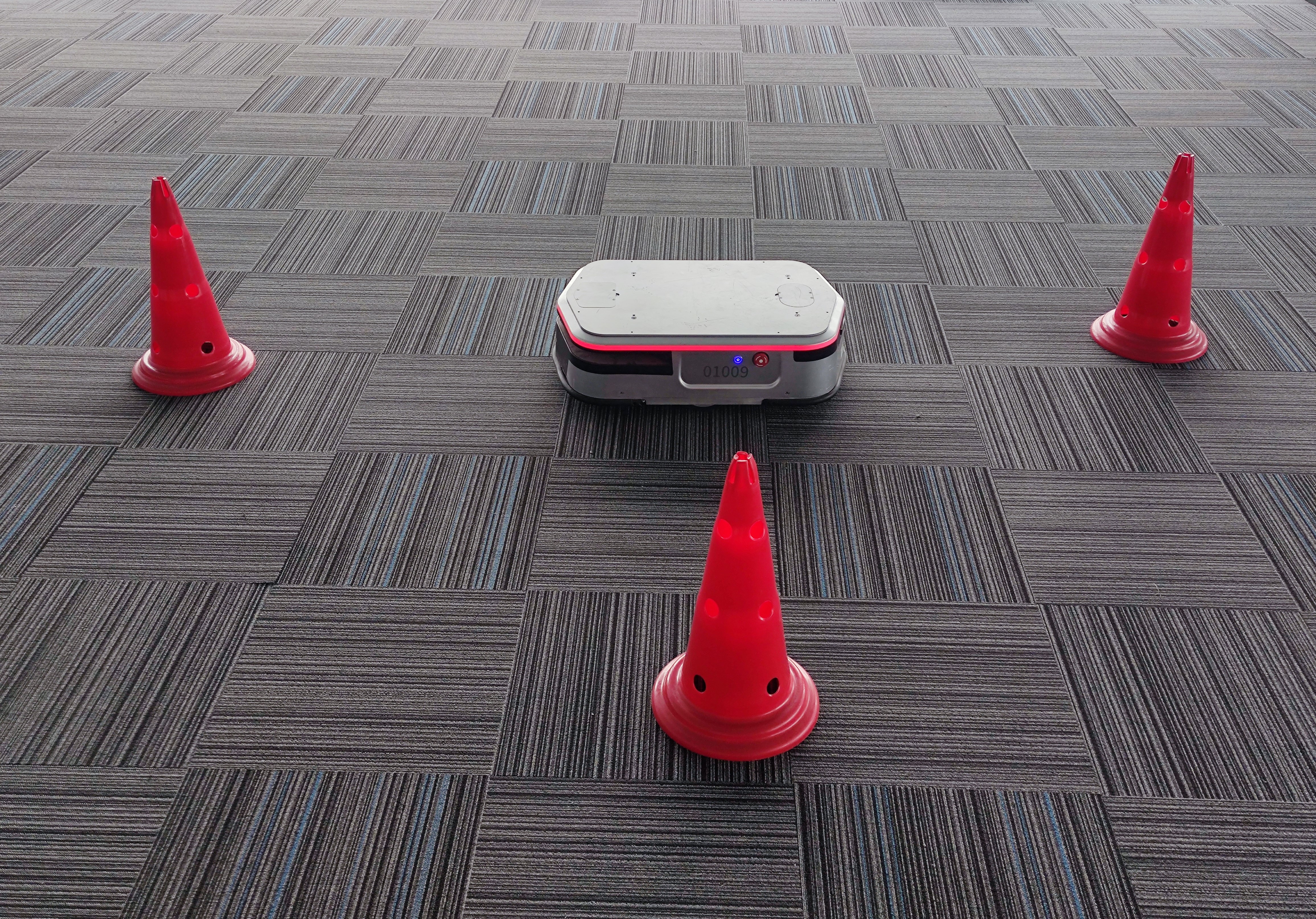}}
\caption{(a) The warehouse vehicle. (b) Experiment field. The three red traffic cones serve as landmarks for positioning.}
\label{fig.exp platform}
\end{figure}

In this experiment, we utilized a differential-drive vehicle equipped with a Lidar sensor for environment perception, as depicted in Fig. \ref{fig.exp platform}. The vehicle state is represented by $x=\left[\begin{matrix}p_x &p_y&\theta\end{matrix}\right]^\top$, including its 2D position ($p_x$ and $p_y$) and orientation angle ($\theta$). The longitudinal velocity $v$ and yaw rate $\omega$ are the control inputs \cite{elfring2021particle}. The Lidar sensor boasts a 240-degree detection range and 0.33-degree resolution. Throughout the experiment, manual control was employed to guide the vehicle's movement.

\begin{figure}[!htb]
\centering 
% \captionsetup[subfigure]{justification=centering}
\subfloat[]{\includegraphics[width=0.42\linewidth]{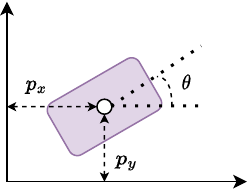}}\hspace{1mm} 
\subfloat[]{\includegraphics[width=0.48\linewidth]{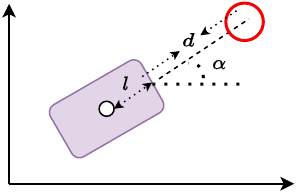}}
\caption{(a) The diagram of the vehicle's states containing the 2D position of the vehicle $p_x$ and $p_y$, and the orientation angle $\theta$. (b) The diagram of the measurement model. Note that the red circle represents the red traffic cone.}
\label{fig.robot system diagram}
\end{figure}

The dynamics of the vehicle can be modeled as:
\begin{equation}\label{eq.vehicle dynamics}
\begin{aligned}
    \begin{bmatrix}
        p_{x,t+1}\\
        p_{y,t+1}\\
        \theta_{t+1}
\end{bmatrix}=\begin{bmatrix}
        p_{x,t}\\
        p_{y,t}\\
        \theta_t
    \end{bmatrix} + \begin{bmatrix}
        v_t \cdot \cos\theta_t\\
        v_t \cdot \sin\theta_t\\
        \omega_t
    \end{bmatrix}\cdot\Delta t + \xi_t,
\end{aligned}
\end{equation}
where $\Delta t=0.0667s$ is the sample period, $v_t$ is the longitudinal velocity, $\omega_t$ is the yaw rate, and $\xi_t$ is the process noise. In addition, the vehicle relies on three red traffic cones with predetermined positions as landmarks on the map to estimate its own location. The observation model is given by:
\begin{equation}\label{eq.vehicle measurement equation}
    y_t = [d^1_t, d^2_t, d^3_t, \alpha^1_t, \alpha^2_t, \alpha^3_t]^\top + \zeta_t,
\end{equation}
where $d^i$ and $\alpha^i$ ($i=1, 2, 3$) are defined as relative distance and orientation angle between the vehicle and the traffic cones respectively:
\begin{equation}
\begin{aligned}
    d^i &= \|(p^{tc,i}_{x}-p_x-l\cos\theta, p^{tc,i}_{y}-p_y-l \sin\theta)\|_2,\\
    \alpha^i &= \arctan\left(\frac{p^{tc,i}_{y}-p_y-l \sin\theta}{p^{tc,i}_{x}-p_x-l\cos\theta}\right) - \theta.
\end{aligned}
\end{equation}
Here $(p^{tc,i}{x}, p^{tc,i}{y})$ is the position of the $i_{\text{th}}$ traffic cone, and $l$ represents the longitudinal installation offset of the Lidar with respect to the robot center. The system model is illustrated in Fig. \ref{fig.robot system diagram}.

We use an auxiliary high-definition indoor SLAM system to acquire highly precise location estimates for the vehicle, with a localization error less than $10 mm$. The output of the SLAM system is treated as the ground truth states for evaluation. The detailed parameters about the experimental setup are given in Table \ref{tab.parameters of the experiment}.
\begin{table}
\begin{center}
\caption{Parameters of the experiments.}
\label{tab.parameters of the experiment}
\begin{tabular}{| c | c | c |}
\hline
Symbol & Description & Value\\
\hline
$(p^{lm}_{x,1}, p^{lm}_{y,1})$ & $1\text{st}$ traffic cone's position    & (1.05m, -2.69m)   \\
$(p^{lm}_{x,2}, p^{lm}_{y,2})$ & $2\text{nd}$ traffic cone's position     & (4.07m, -1.75m)  \\
$(p^{lm}_{x,3}, p^{lm}_{y,3})$ & $3\text{rd}$ traffic cone's position   & (6.02m, -3.32m)   \\
$l$ &Lidar longitude offset& 0.329m\\
\hline 
\end{tabular}
\end{center}
\end{table}
\subsection{Data Pre-processing and Experimental Results}
The experiment begins with the gathering of raw data, including ground truth states, control inputs, and the Lidar point cloud. Subsequently, we process the Lidar point cloud to extract measurements, which represent the relative distance and orientation angle between the vehicle and the traffic cones. The assembled dataset incorporates trajectories totaling 11 minutes in length.

\begin{figure}[!htb]
\centering
\includegraphics[width=0.98\linewidth]{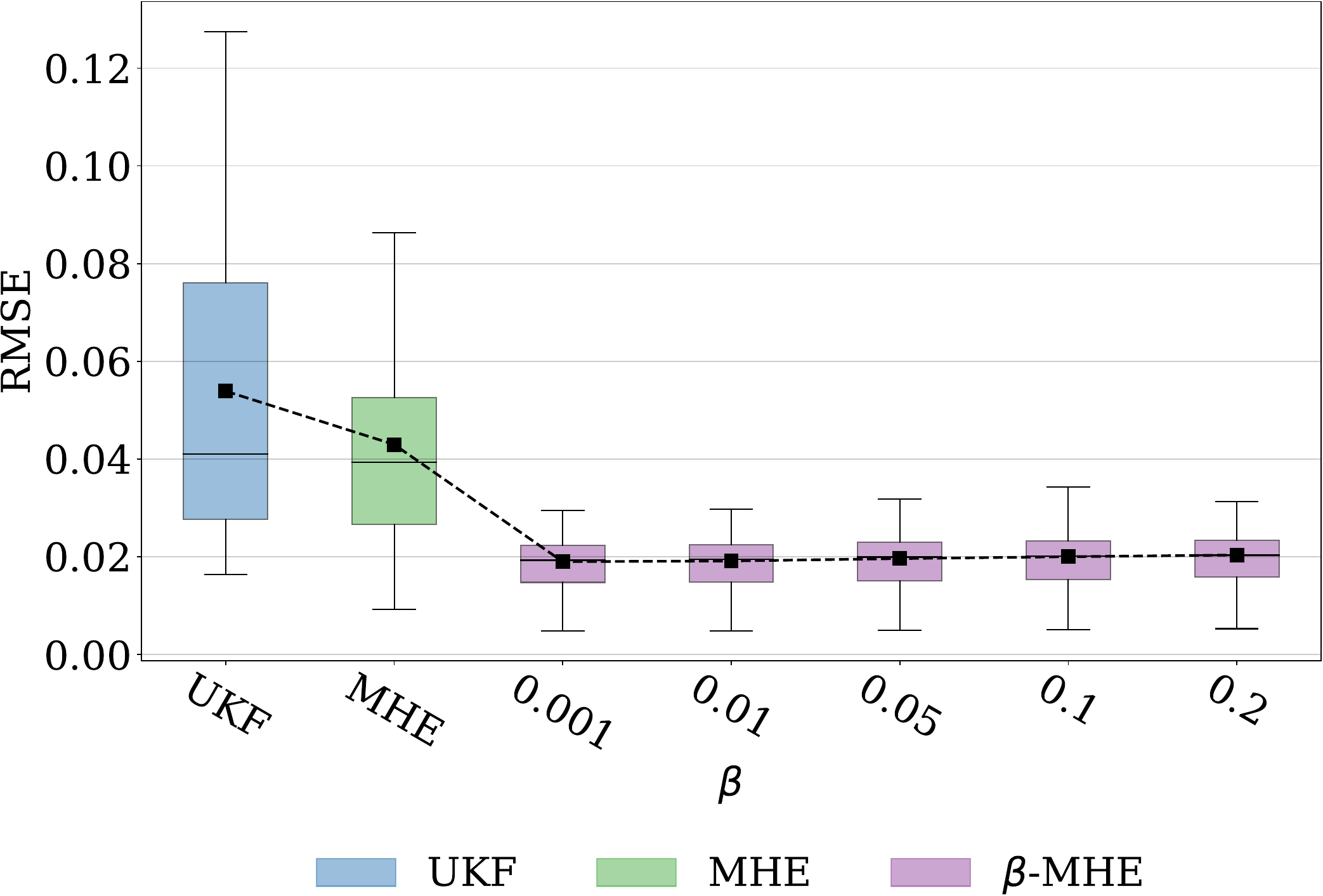}
\caption{Box plot of RMSE for UKF, MHE and $\beta$-MHE with different values of $\beta$ when contamination probability $p_c=0.05$. The black square `` $\blacksquare$ " represents the average RMSE.}\label{fig.robot boxplot}
\end{figure}

\begin{figure}[!htb]
\centering
\includegraphics[width=0.98\linewidth]{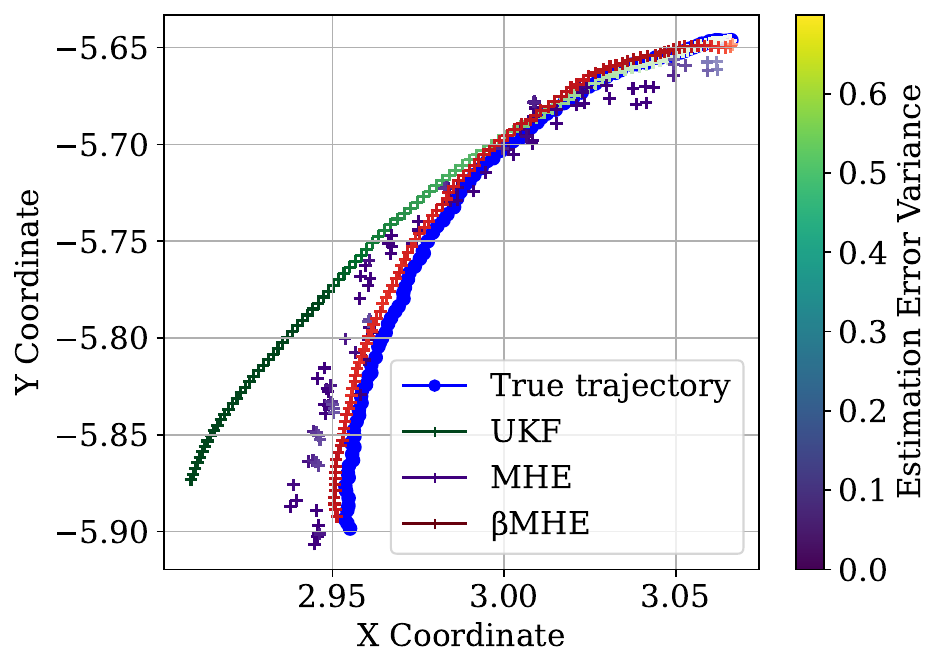}
\caption{Robot trajectories with estimation error variance for UKF, MHE and $\beta$-MHE with $\beta=0.1$.}\label{fig.robot traj}
\end{figure}

Following this, we undertake the noise identification process, where we employ the ground truth data and measurements to obtain the parameters of the noise distribution. We operate under the assumption that both process and observation noises adhere to a Gaussian distribution, with each component being mutually independent. Specifically, the variance matrix of the Gaussian distribution manifests as a diagonal matrix. The noises can be represented as
\begin{equation}
\begin{aligned}\label{eq.noise distribution}
\xi_t \sim \mathcal{N}
\left(\mu_{\xi}
, \Sigma_{\xi}\right),
\zeta_t \sim \mathcal{N}
\left(\mu_{\zeta}
, \Sigma_{\zeta}\right).
\end{aligned}
\end{equation}
For the acquired samples of process noise and measurement noise, we utilize \eqref{eq.vechile noises} to implement a maximum likelihood estimation to identify the mean and variance of the distributions. 
%Fig. \ref{fig.noise dist}.
\begin{equation}\label{eq.vechile noises}
\begin{aligned}
\xi_t  &=\begin{bmatrix}
  p_{x,t+1}\\
        p_{y,t+1}\\
        \theta_{t+1}
\end{bmatrix} - \begin{bmatrix}
        p_{x,t}\\
        p_{y,t}\\
        \theta_t
    \end{bmatrix} - \begin{bmatrix}
        v_t \cdot \cos\theta_t\\
        v_t \cdot \sin\theta_t\\
        \omega_t
    \end{bmatrix}\cdot\Delta t,\\
\zeta_t &= y_t - [d^1_t, d^2_t, d^3_t, \alpha^1_t, \alpha^2_t, \alpha^3_t]^\top.
\end{aligned}    
\end{equation}
The identified parameters of the noise are:
\begin{equation}\nonumber
\begin{aligned}
\mu_{\xi} &= \left[\begin{matrix}
-0.00017&
-0.00020&
-0.00056  
\end{matrix}
\right]^\top,
\\
\Sigma_{\xi} &= \text{diag}\Big(
\left[
\begin{matrix}
0.0034&
0.0056&
0.0041
\end{matrix}
\right]
\Big),
\\
\mu_{\zeta} &= \left[\begin{matrix}
-0.0312&
-0.0581&
-0.0557&
0.0053&
0.0059&
0.0125
\end{matrix}
\right]^\top,
\\
\Sigma_{\zeta} &= \text{diag}\Big(
\left[
\begin{matrix}
0.0238&
0.0284&
0.0259&
0.0107&
0.0094&
0.0118
\end{matrix}
\right]
\Big).
\end{aligned}    
\end{equation}
% \subsection{Experimental Results}
In the experiment, we simulate a scenario where the Lidar sensor might occasionally penetrate objects, thereby generating an output equivalent to the maximum Lidar range of 20m. Assuming that such an outlier is encountered with a probability of $p_c = 0.01$, we segment the dataset into 50 sub-trajectories, each lasting 6.67 seconds (equivalent to 100 time steps). The experimental results are presented in Fig. \ref{fig.robot boxplot} and Fig. \ref{fig.robot traj}, which provide a comparative analysis of the performance metrics of the UKF, MHE, and $\beta$-MHE under different $\beta$ values. The experimental results indicate that $\beta$-MHE  yields superior performance than both MHE and UKF across various $\beta$ values.

\section{Conclusion}
\label{sec.VIIII}
In this paper, we proposed a robust Bayesian inference framework for MHE that can maintain its accuracy in the presence of measurement outliers. By introducing a robust divergence measure, which assigns greater weight to inliers, our method mitigates the impact of outliers and significantly improves the robustness of the MHE method. We derived the analytical form of the influence functions for MHE methods. We demonstrated that, for linear Gaussian systems, the influence function of our proposed method is bounded, highlighting its robustness. Moreover, we performed stability analysis and shown that our proposed method exhibits robust asymptotic stability under certain conditions. 
Future studies will focus on developing techniques for auto-tuning parameters and applications to large-scale systems.

\bibliographystyle{plain}        % Include this if you use bibtex 
\bibliography{autosam}           % and a bib file to produce the 

\appendix

\section{Proof of Theorem \ref{theorem.IF of MHE}}\label{appendix.theorem 1}

\begin{proof}
Consider that the measurement model is contaminated by outliers, and the empirical probability satisfies
\begin{equation}\label{eq.contaminated emperical distribution}
p_\text{em}^t(y)=(1-\epsilon)\delta(y-y_t)+\epsilon\delta(y-z).    
\end{equation}
Substituting \eqref{eq.contaminated emperical distribution} for $p^t_\text{true}(y)$ in \eqref{eq.modified beta divergence}, the loss function $\ell^{\beta}(x_t, y_t)$ defined in \eqref{eq.beta loss} is converted to
\begin{equation}\label{eq.contaminated beta loss}\nonumber
\begin{aligned}
\ell^{\beta,z,\epsilon}(x_t,y_t)
&=- \frac{(1-\epsilon)(\beta+1)}{\beta}g(y_t|x_t) ^{\beta} \\ &- \frac{\epsilon(\beta+1)}{\beta}g(z|x_t) ^{\beta}
+\int{g(y|x_t)^{\beta+1}\mathrm{d}y}.
\end{aligned}
\end{equation}
If we replace $G^{\beta}(y_t|x_t)$ defined in \eqref{eq.general beta likelihood} with $G^{\beta,z,\epsilon}(y_t|x_t) :=\exp{\left(-\ell^{\beta,z,\epsilon}(x_t,y_t)\right)}$, the state estimate with contaminated data can be obtained by
\begin{equation}\nonumber
\hat{x}_{t-T:t}^{z, \epsilon} = \arg\min_{x_{t-T:t}}\left\{J^{\beta,z,\epsilon}(x_{t-T:t})\right\},  
\end{equation}
where
\begin{equation}\nonumber
\begin{aligned}
&J^{\beta,z,\epsilon}(x_{t-T:t}) \\
&=\Gamma(x_{t-T})+\sum\limits_{i=t-T+1}^{t} \left\{k(x_{i},x_{i-1})+\ell^{\beta,z,\epsilon}(x_i,y_i)\right\}.
\end{aligned}
\end{equation}
Here $\Gamma(x_{t-T})$ satisfies \eqref{eq.mhe-1}, and $k(x_t,x_{t-1})$ satisfies \eqref{eq.mhe-2}.
By the assumption of the differentiability of $J^{\beta,z,\epsilon}(x_{t-T:t})$, we obtain
\begin{equation}\label{eq.necessary condition}
\begin{aligned}
\left.\frac{\partial{\left\{J^{\beta,z ,\epsilon}(x_{t-T:t})\right\}}}{\partial{x_{t-T:t}}}\right|_{x_{t-T:t}=\hat{x}_{t-T:t}^{z, \epsilon}} = 0.
\end{aligned}
\end{equation}
According to the definition,
\begin{equation}\nonumber
\mathbb{IF}(z, \hat{x}_{t-T: t}, G^{\beta}(y_t|x_t)) = \left.\frac{\partial{\hat{x}_{t-T:t}^{z, \epsilon}}}{\partial{\epsilon}}\right|_{\epsilon=0}.    
\end{equation}
Taking the derivative of the both sides of \eqref{eq.necessary condition} with respect to $\epsilon$ at $\epsilon = 0$ and defining $\mathscr{K}^{\beta, z}(x_t,y_t)=\left.\frac{\partial{\ell^{\beta,z,\epsilon}(x_t,y_t)}}{\partial{\epsilon}}\right|_{\epsilon=0}$, we have
\begin{equation}\nonumber
\begin{aligned}
&\left.\frac{\partial^2{\left[J(x_{t-T:t})\right]}}{\partial{x_{t-T:t}^2}}\right|_{{x_{t-T:t}=\hat{x}_{t-T:t}}} \cdot \mathbb{IF}(z, \hat{x}_{t-T: t}, G^{\beta}(y_t|x_t))\\
&+\sum\limits_{i=t-T+1}^t\left.\frac{\partial{\left[\mathscr{K}^{\beta, z}(x_i,y_i)\right]}}{\partial{x_{t-T:t}}}\right|_{{x_{t-T:t}=\hat{x}_{t-T:t}}} = 0.
\end{aligned}
\end{equation}
Here, $J(x_{t-T:t})$ is defined by \eqref{eq.mhe-b}\eqref{eq.mhe-1}\eqref{eq.mhe-2}\eqref{eq.beta-mhe-3}. Therefore, \eqref{eq. IF of beta MHE} holds given the Assumption \ref{assump.diffenertiable} that the Hessian matrix $M_1(\hat{x}_{t-T: t})$ is nonsingular.
\end{proof}

\section{Proof of Theorem 2}\label{appendix.theorem 2}
\begin{proof}
First, we will prove that the gross error sensitivity of MHE is infinite.
We observe that
\begin{equation}\label{eq.bound for MHE}\nonumber
\Vert M_2(z, \hat{x}_{t-T: t}) \Vert \leq \Vert M_1(\hat{x}_{t-T: t}) \Vert_{\mathcal{F}} \cdot \Vert \mathbb{IF}(z, \hat{x}_{t-T: t}, g(y_t|x_t)) \Vert.
\end{equation}
For MHE, we have
\begin{equation}\label{eq.M2 matrix for MHE a}
\sum\limits_{j=t-T+1}^t\left.\frac{\partial{\left[\mathscr{K}^{z}(x_j,y_j)\right]}}{\partial{x_{t-T}}}\right|_{x_{t-T}=\hat{x}_{t-T}} = 0,
\end{equation}
and
\begin{equation}\label{eq.M2 matrix for MHE b}
\begin{aligned}
&\sum\limits_{j=t-T+1}^t\left.\frac{\partial{\left[\mathscr{K}^{z}(x_j,y_j)\right]}}{\partial{x_{t-i}}}\right|_{x_{t-i}=\hat{x}_{t-i}} = C^{\top}R^{-1}(y_{t-i}-C\hat{x}_{t-i})\\
&-C^{\top}R^{-1}(z-C\hat{x}_{t-i}),\;\;i \in \mathbb{Z}_{[0, T-1]}.
\end{aligned}
\end{equation}
Thus $\sup_z \Vert M_2(z, \hat{x}_{t-T: t}) \Vert = \infty$, which leads to
\begin{equation}\nonumber
\begin{aligned}
\gamma(\hat{x}_{t-T: t}, g(y_t|x_t))&=
\sup_z{\Vert \mathbb{IF}(z, \hat{x}_{t-T: t}, g(y_t|x_t)) \Vert}\\
&\geq \frac{\sup_z \Vert M_2(z, \hat{x}_{t-T: t}) \Vert}{\Vert M_1(\hat{x}_{t-T: t}) \Vert_{\mathcal{F}}}\\&
=\infty.
\end{aligned}
\end{equation}
In the next step, we will prove that the gross error sensitivity of the $\beta$-MHE is bounded. We find
\begin{equation}\nonumber
\begin{aligned}
\Vert \mathbb{IF}(z, \hat{x}_{t-T: t}, G^{\beta}(y_t|x_t)) \Vert \leq &\Vert M_1^{-1}(\hat{x}_{t-T: t}) \Vert_{\mathcal{F}} \\
\cdot &\Vert M_2(z, \hat{x}_{t-T: t}) \Vert.
\end{aligned}
\end{equation}
Similar to \eqref{eq.M2 matrix for MHE a} and \eqref{eq.M2 matrix for MHE b}, we have
\begin{equation}\label{eq.M2 matrix for beta MHE}\nonumber
\sum\limits_{j=t-T+1}^t\left.\frac{\partial{\left[\mathscr{K}^{\beta,z}(x_j,y_j)\right]}}{\partial{x_{t-T}}}\right|_{x_{t-T}=\hat{x}_{t-T}} = 0,
\end{equation}
and
\begin{equation}\nonumber
\begin{aligned}
&\sum\limits_{j=t-T+1}^t\left.\frac{\partial{\left[\mathscr{K}^{\beta,z}(x_j,y_j)\right]}}{\partial{x_{t-i}}}\right|_{x_{t-i}=\hat{x}_{t-i}}
\\&=\frac{(\beta+1)C^{\top}R^{{-}1}}{\sqrt{(2\pi)^{\beta m}|R|^{\beta}}}(y_{t-i}-C\hat{x}_{t-i})e^{-\frac{\beta}{2}\Vert y_{t-i}-C\hat{x}_{t-i} \Vert^2_{{R}^{-1}}}\\
&-\frac{(\beta+1)C^{\top}R^{-1}}{\sqrt{(2\pi)^{\beta m}|R|^{\beta}}}(z-C\hat{x}_{t-i})e^{-\frac{\beta}{2}\Vert z-C\hat{x}_{t-i} \Vert^2_{{R}^{-1}}},\\
&\forall i\in \mathbb{Z}_{[0, T-1]}.
\end{aligned}
\end{equation}
Define the function $\rho(\hat{x}_t,z)$ as in \eqref{eq.definition of rho} and 
% $\rho(\hat{x}_t,z)$ satisfies
% \begin{equation}\nonumber
% \begin{aligned}
% \frac{\partial \Vert \rho(\hat{x}_t,z) \Vert^2 }{\partial{z}} &=2\frac{\left[\mathbb{I}_{m \times m}-\beta R^{-1}(z-C\hat{x}_t)(z-C\hat{x}_t)^{\top}\right]}{\sqrt{(2\pi)^{\beta m}|R|^{\beta}}} \\
% &\cdot R^{-1}C \cdot e^{-\frac{\beta}{2}\Vert z-C\hat{x}_{t} \Vert^2_{{R}^{-1}}}\cdot \rho(\hat{x}_t,z).
% \end{aligned}
% \end{equation}
% We suppose $R=L_RL_R^\top$ by Cholesky decomposition. Because $z = C\hat{x}_t+\frac{L_R}{\sqrt{\beta}}$, $z = C\hat{x}_t-\frac{L_R}{\sqrt{\beta}}$ and $z = C\hat{x}_t$ are the potential
% extreme points 
we observe that 
$\lim_{z \to \infty} \rho(\hat{x}_t,z) = 0$.
Because $\Vert \rho(\hat{x}_t, z) \Vert^2$ is continuous with respect to $z$, it is bounded, i.e.,
\begin{equation}\nonumber
\begin{aligned}
\Vert \rho(\hat{x}_t,z) \Vert \leq & \rho_{\text{max}},\;\;\forall z.
\end{aligned} 
\end{equation}
Because
\begin{equation}\nonumber
\begin{aligned}
&\Vert M_2(z,\hat{x}_{t-T: t}) \Vert^2 \\
&=\sum\limits_{i=0}^{T-1}\left\Vert \sum\limits_{j=t-T+1}^t\left.\frac{\partial{\left[\mathcal{K}^{\beta,z}(x_j,y_j)\right]}}{\partial{x_{t-i}}}\right|_{x_{t-i}=\hat{x}_{t-i}}\right\Vert^2\\
&\leq 4T\cdot\rho_{\text{max}}^2,
\end{aligned}
\end{equation}
we get
\begin{equation}\label{eq.bound of IF}
\begin{aligned}
\Vert \mathbb{IF}(z, \hat{x}_{t-T: t}, G^{\beta}(y_t|x_t)) \Vert &\leq 2\sqrt{T} \cdot\Vert M_1^{-1}(\hat{x}_{t-T: t}) \Vert_{\mathcal{F}}\cdot \rho_{\text{max}}.  
\end{aligned}
\end{equation}
Taking the supremum on both sides of \eqref{eq.bound of IF} results in \eqref{eq.bound for beta MHE}, thereby completing the proof.
\end{proof}

\section{Proof of Theorem 3}\label{appendix.Theorem 3}
In order to prove the robust stability of MHE, we first need a proposition that bounds the estimator error within the MHE problem in terms of the initial state estimate error and the norm of disturbances. 
\begin{proposition}\label{proposition.FIE RAS}
Under Assumption \ref{assump.well-posed}, for $\forall t \in \mathbb{Z}_{[0, T]}$, we have
{\small
\begin{equation}\label{eq.FIE RAS}
\begin{aligned}
&\|\hat{x}_{t|t} - x_{t}\|
\\
\leq &\gamma_x\left(
2\sqrt{3} \cdot \|{x}_{0} - \hat{x}_{0|0}\|, t \right) \oplus \gamma_x \left(2\sqrt {\frac{3T}{\mu} \cdot \Bar{\gamma}_{\xi} \left( \| \xi \|_{0:t-1} \right) }, t \right)
\\
\oplus &\gamma_x\left(
2\sqrt{\frac{3 
T}{\mu} \cdot \Bar{\gamma}_{\zeta} \left( \| \zeta \|_{1:t} \right)
}, t \right) 
\oplus
\gamma_{\xi} \left( 2\underline{\gamma}_{\xi}^{-1}
\Big(
3\mu \cdot \|{x}_{0} - \hat{x}_{0|0}\|^2 
\Big) \right)
\\
\oplus&
\gamma_{\xi} \left( 2\underline{\gamma}_{\xi}^{-1}
\Big(
3T \cdot \Bar{\gamma}_{\xi} \left( \| \xi \|_{0:t-1} \right) \Big) \right) \oplus
\gamma_{\xi} \left( 2\underline{\gamma}_{\xi}^{-1}
\Big(
3T \cdot \Bar{\gamma}_{\zeta} \left( \| \zeta \|_{1:t} \right) \Big) \right)
\\
\oplus& \gamma_{\zeta}\left( 2\underline{\gamma}_{\zeta}^{-1}
\Big(
3\mu \cdot \|{x}_{0} - \hat{x}_{0|0}\|^2 
\Big) \right) 
\oplus
\gamma_{\zeta}\left(2\underline{\gamma}_{\zeta}^{-1}
\Big(
3T \cdot \Bar{\gamma}_{\xi} \left( \| \xi \|_{0:t-1} \right) \Big) \right)
\\
\oplus& \gamma_{\zeta}\left(2\underline{\gamma}_{\zeta}^{-1} \Big(
3T \cdot \Bar{\gamma}_{\zeta} \left( \| \zeta \|_{1:t} \right) \Big) \right).
\end{aligned}
\end{equation}}
\end{proposition}
\begin{proof}
By Assumption \ref{assump.well-posed}, the MHE problem exists a solution $\hat{x}_{0|t}, \hat{\xi}_{0:t-1|t} , \hat{\zeta}_{1:t|t} $. For $\forall t \in \mathbb{Z}_{[0, T]}$, we have
\begin{equation}\nonumber
\begin{aligned}
&\mu \cdot \|\hat{x}_{0|t} - \hat{x}_{0|0}\|^2 \oplus \underline{\gamma}_{\xi} \left( \| \hat{\xi} \|_{0:t-1} \right) \oplus \underline{\gamma}_{\zeta} \left( \| \hat{\zeta} \|_{1:t} \right)
\\
\leq &\mu \cdot \|\hat{x}_{0|t} - \hat{x}_{0|0}\|^2 + \underline{\gamma}_{\xi} \left( \| \hat{\xi} \|_{0:t-1} \right) + \underline{\gamma}_{\zeta} \left( \| \hat{\zeta} \|_{1:t} \right)
\\
\leq & J(\hat{x}_{0|t} , \hat{\xi}_{0:t-1|t} )
\\
\leq & J(\hat{x}_{0|t} , {\xi}_{0:t-1|t} )
\\
\leq & 3\mu \cdot \|{x}_{0} - \hat{x}_{0|0}\|^2 \oplus 3T \cdot \Bar{\gamma}_{\xi} \left( \| \xi \|_{0:t-1} \right) \oplus 3T \cdot \Bar{\gamma}_{\zeta} \left( \| \zeta \|_{1:t} \right).
\end{aligned}
\end{equation}
Thus, we obtain the following bounds
{\small
\begin{subequations}\label{eq.bound1}
\begin{align}
&\|\hat{x}_{0|t} - \hat{x}_{0|0}\| \nonumber
\\ 
&\leq 
\frac{\sqrt{3}}{\sqrt{\mu}} \Big(
\sqrt{
\mu \cdot \|{x}_{0} - \hat{x}_{0|0}\|^2 \oplus T \cdot \Bar{\gamma}_{\xi} \left( \| \xi \|_{0:t-1} \right) \oplus 
T \cdot \Bar{\gamma}_{\zeta} \left( \| \zeta \|_{1:t} \right)
}
\Big)
\nonumber \\
& =
\sqrt{3} \cdot \|{x}_{0} - \hat{x}_{0|0}\| \oplus \sqrt {\frac{3T}{\mu} \cdot \Bar{\gamma}_{\xi} \left( \| \xi \|_{0:t-1} \right) } \oplus \sqrt{\frac{3 
T}{\mu} \cdot \Bar{\gamma}_{\zeta} \left( \| \zeta \|_{1:t} \right)
},
\\
&\| \hat{\xi} \|_{0:t-1} \nonumber
\\
& \leq 
\underline{\gamma}_{\xi}^{-1}
\Big(
3
\Big(
\mu \cdot \|{x}_{0} - \hat{x}_{0|0}\|^2 \oplus T \cdot \Bar{\gamma}_{\xi} \left( \| \xi \|_{0:t-1} \right) \oplus 
T \cdot \Bar{\gamma}_{\zeta} \left( \| \zeta \|_{1:t} \right)
\Big)\Big)
\nonumber \\
&= 
\underline{\gamma}_{\xi}^{-1}
\Big(
3\mu \cdot \|{x}_{0} - \hat{x}_{0|0}\|^2 
\Big)
\oplus
\underline{\gamma}_{\xi}^{-1}
\Big(
3T \cdot \Bar{\gamma}_{\xi} \left( \| \xi \|_{0:t-1} \right) \Big)
\nonumber
\\
& \oplus
\underline{\gamma}_{\xi}^{-1} \Big(
3T \cdot \Bar{\gamma}_{\zeta} \left( \| \zeta \|_{1:t} \right) \Big),
\\
&\| \hat{\zeta} \|_{1:t} \nonumber
\\
& \leq 
\underline{\gamma}_{\zeta}^{-1}
\Big(
3
\Big(
\mu \cdot \|{x}_{0} - \hat{x}_{0|0}\|^2 \oplus T \cdot \Bar{\gamma}_{\xi} \left( \| \xi \|_{0:t-1} \right) \oplus 
T \cdot \Bar{\gamma}_{\zeta} \left( \| \zeta \|_{1:t} \right)
\Big)\Big)
\nonumber \\
&= 
\underline{\gamma}_{\zeta}^{-1}
\Big(
3\mu \cdot \|{x}_{0} - \hat{x}_{0|0}\|^2 
\Big)
\oplus
\underline{\gamma}_{\zeta}^{-1}
\Big(
3T \cdot \Bar{\gamma}_{\xi} \left( \| \xi \|_{0:t-1} \right) \Big)
\nonumber
\\
& \oplus
\underline{\gamma}_{\zeta}^{-1} \Big(
3T \cdot \Bar{\gamma}_{\zeta} \left( \| \zeta \|_{1:t} \right) \Big).
\end{align}    
\end{subequations}}
From the system’s i-IOSS bound in \eqref{eq.IOSS}, we have
\begin{equation}
\label{eq.bound2}
\begin{aligned}
&\|\hat{x}_{t|t} - x_{t}\|
\\
\leq
&\gamma_x(
\|\hat{x}_{0|t} - x_{0}\|, t)  
\oplus  \gamma_{\xi} \left(\| {\xi} - \hat{\xi} 
\|_{0:t-1} \right) 
\oplus  \gamma_{\zeta} \left(\| \zeta - \hat \zeta 
\|_{1:t} \right)
\\
\leq
&\gamma_x(
\|\hat{x}_{0|t} - \hat{x}_{0|0} + \hat{x}_{0|0} - x_{0}\|, t) \\ 
\oplus & \gamma_{\xi} (\left\| {\xi}\right\|_{0:t-1} + | \hat{\xi} \|_{0:t-1} ) 
\oplus  \gamma_{\zeta} \left(\| {\zeta}\|_{1:t} + \| \hat{\zeta} 
\|_{1:t}\right)
\\
\leq
&\gamma_x(
2\|\hat{x}_{0|t} - \hat{x}_{0|0}\| \oplus 2| \hat{x}_{0|0} - x_{0}|, t)
\\ 
\oplus & \gamma_{\xi} (2\left \| {\xi}\right\|_{0:t-1} \oplus 2 \| \hat{\xi} 
\|_{0:t-1} ) 
\oplus  \gamma_{\zeta} (2\| {\zeta}\|_{1:t} \oplus 2\| \hat{\zeta} 
\|_{1:t}).
\end{aligned}
\end{equation}

Substituting \eqref{eq.bound1} into \eqref{eq.bound2}, we have
{\small
\begin{equation}\nonumber
\begin{aligned}
&\|\hat{x}_{t|t} - x_{t}\|
\\
\leq & \gamma_x(
2\|\hat{x}_{0|t} - \hat{x}_{0|0}\| \oplus 2\| \hat{x}_{0|0} - x_{0}\|, t)
\\ 
\oplus & \gamma_{\xi} (2\left \| {\xi}\right\|_{0:t-1} \oplus 2 \| \hat{\xi} 
\|_{0:t-1} ) 
\oplus  \gamma_{\zeta} (2\| {\zeta}\|_{1:t} \oplus 2\| \hat{\zeta} 
\|_{1:t})
\\
\leq
&\gamma_x(
2\| \hat{x}_{0|0} - x_{0}\|, t)
\oplus \gamma_{\xi} (2\left \| {\xi}\right\|_{0:t-1}) 
\oplus  \gamma_{\zeta} (2\left\| {\zeta}\right\|_{1:t})
\\
\oplus &\gamma_x\left(
2\sqrt{3} \cdot \|{x}_{0} - \hat{x}_{0|0}\|, t \right) \oplus \gamma_x \left(2\sqrt {\frac{3T}{\mu} \cdot \Bar{\gamma}_{\xi} \left( \| \xi \|_{0:t-1} \right) }, t \right)
\\
\oplus &\gamma_x\left(
2\sqrt{\frac{3 
T}{\mu} \cdot \Bar{\gamma}_{\zeta} \left( \| \zeta \|_{1:t} \right)
}, t \right) 
\oplus
\gamma_{\xi} \left( 2\underline{\gamma}_{\xi}^{-1}
\Big(
3\mu \cdot \|{x}_{0} - \hat{x}_{0|0}\|^2 
\Big) \right)
\\
\oplus&
\gamma_{\xi} \left( 2\underline{\gamma}_{\xi}^{-1}
\Big(
3T \cdot \Bar{\gamma}_{\xi} \left( \| \xi \|_{0:t-1} \right) \Big) \right) \oplus
\gamma_{\xi} \left( 2\underline{\gamma}_{\xi}^{-1}
\Big(
3T \cdot \Bar{\gamma}_{\zeta} \left( \| \zeta \|_{1:t} \right) \Big) \right)
\\
\oplus& \gamma_{\zeta}\left( 2\underline{\gamma}_{\zeta}^{-1}
\Big(
3\mu \cdot \|{x}_{0} - \hat{x}_{0|0}\|^2 
\Big) \right) 
\oplus
\gamma_{\zeta}\left(2\underline{\gamma}_{\zeta}^{-1}
\Big(
3T \cdot \Bar{\gamma}_{\xi} \left( \| \xi \|_{0:t-1} \right) \Big) \right)
\\
\oplus& \gamma_{\zeta}\left(2\underline{\gamma}_{\zeta}^{-1} \Big(
3T \cdot \Bar{\gamma}_{\zeta} \left( \| \zeta \|_{1:t} \right) \Big) \right).
\end{aligned}
\end{equation}}

Combine with the fact that
\begin{equation}\nonumber
\begin{aligned}
\gamma_x(
2\| \hat{x}_{0|0} - x_{0}\|, t) &\leq
\gamma_x\left(
2\sqrt{3} \cdot \|{x}_{0} - \hat{x}_{0|0}\|, t \right),
\\
\gamma_{\xi} (2\left\| {\xi}\right\|_{0:t-1}) & \leq
\gamma_{\xi} \left( 2\underline{\gamma}_{\xi}^{-1}
\Big(
3T \cdot \Bar{\gamma}_{\xi} \left( \| \xi \|_{0:t-1} \right) \Big) \right),
\\
\gamma_{\zeta} (2\left \| {\zeta}\right\|_{1:t}) &
\leq
\gamma_{\zeta}\left(2\underline{\gamma}_{\zeta}^{-1} \Big(
3T \cdot \Bar{\gamma}_{\zeta} \left( \| \zeta \|_{1:t} \right) \Big) \right),
\end{aligned}
\end{equation}
we can finally obtain \eqref{eq.FIE RAS}.
\end{proof}

Then, we give the proof of Theorem \ref{theorem.stability}.

\begin{proof}
By Proposition \ref{proposition.FIE RAS}, we have {\small \begin{equation}\label{eq.FIE RAS 1}
\begin{aligned}
&\|\hat{x}_{t+T|t+T} - x_{t+T}\|
\\
\leq &\gamma_x\left(
2\sqrt{3} \cdot \|{x}_{t} - \hat{x}_{t|t}\|, T \right) \oplus \gamma_x \left(2\sqrt {\frac{3T}{\mu} \cdot \Bar{\gamma}_{\xi} \left( \| \xi \|_{t:t+T-1} \right) }, T \right)
\\
\oplus &\gamma_x\left(
2\sqrt{\frac{3 
T}{\mu} \cdot \Bar{\gamma}_{\zeta} \left( \| \zeta \|_{t+1:t+T} \right)
}, T \right) 
\oplus
\gamma_{\xi} \left( 2\underline{\gamma}_{\xi}^{-1}
\Big(
3\mu \cdot \|{x}_{t} - \hat{x}_{t|t}\|^2 
\Big) \right)
\\
\oplus&
\gamma_{\xi} \left( 2\underline{\gamma}_{\xi}^{-1}
\Big(
3T \cdot \Bar{\gamma}_{\xi} \left( \| \xi \|_{t:t+T-1} \right) \Big) \right) \oplus
\gamma_{\xi} \left( 2\underline{\gamma}_{\xi}^{-1}
\Big(
3T \cdot \Bar{\gamma}_{\zeta} \left( \| \zeta \|_{t+1:t+T} \right) \Big) \right)
\\
\oplus& \gamma_{\zeta}\left( 2\underline{\gamma}_{\zeta}^{-1}
\Big(
3\mu \cdot \|{x}_{t} - \hat{x}_{t|t}\|^2 
\Big) \right) 
\oplus
\gamma_{\zeta}\left(2\underline{\gamma}_{\zeta}^{-1}
\Big(
3T \cdot \Bar{\gamma}_{\xi} \left( \| \xi \|_{t:t+T-1} \right) \Big) \right)
\\
\oplus& \gamma_{\zeta}\left(2\underline{\gamma}_{\zeta}^{-1} \Big(
3T \cdot \Bar{\gamma}_{\zeta} \left( \| \zeta \|_{t+1:t+T} \right) \Big) \right).
\end{aligned}
\end{equation}}
Define the $\mathcal{K}$ functions $\alpha_{\xi} (s)$ and $\alpha_{\zeta} (s)$ as:
\begin{equation}\nonumber
\begin{aligned}
\alpha_{\xi} (s) &:= \gamma_x \left(2\sqrt {\frac{3T}{\mu} \cdot \Bar{\gamma}_{\xi} \left( s \right) }, 0 \right) \\
&\oplus \gamma_{\xi} \left( 2\underline{\gamma}_{\xi}^{-1}
\Big(
3T \cdot \Bar{\gamma}_{\xi} \left( s \right) \Big) \right) \oplus \gamma_{\zeta}\left(2\underline{\gamma}_{\zeta}^{-1}
\Big(
3T \cdot \Bar{\gamma}_{\xi} \left( s \right) \Big) \right),
\\
\alpha_{\zeta} (s) &:= \gamma_x \left(2\sqrt {\frac{3T}{\mu} \cdot \Bar{\gamma}_{\zeta} \left( s \right) }, 0 \right) \\
&\oplus \gamma_{\xi} \left( 2\underline{\gamma}_{\xi}^{-1}
\Big(
3T \cdot \Bar{\gamma}_{\zeta} \left( s \right) \Big) \right) \oplus \gamma_{\zeta}\left(2\underline{\gamma}_{\zeta}^{-1}
\Big(
3T \cdot \Bar{\gamma}_{\zeta} \left( s \right) \Big) \right).
\end{aligned}
\end{equation}
Define $\lambda := 2\sqrt{3}\eta < 1$, for $\eta \in (0, \frac{1}{2\sqrt{3}})$,  by Assumption \ref{assump.contraction map}, for $\|x_t - \hat{x}_{t|t}\| \leq \Bar{s}$, there exists $T \geq T_m \geq 0$, such that
\begin{equation}\nonumber
\begin{aligned}
\gamma_x(
2\sqrt{3} \cdot \|{x}_{t} - \hat{x}_{t|t}\|, T)
& \leq
\gamma_x(
2\sqrt{3} \cdot \|{x}_{t} - \hat{x}_{t|t}\|, T_m) 
\\
& \leq 2\sqrt{3} \eta \|{x}_{t} - \hat{x}_{t|t}\| 
\\
&= \lambda |{x}_{t} - \hat{x}_{t|t}|.      
\end{aligned}
\end{equation}
Note that there exists $\delta>0$ such that if $\| \xi \|_{0:\infty}, \| \zeta \|_{0:\infty} \leq \delta$, then $\alpha_{\xi} (\| \xi \|_{0:\infty}), \alpha_{\zeta} (\| \zeta \|_{0:\infty}) \leq \Bar{s}, \forall \Bar{s} >0$.
By Assumption \ref{assump.contraction map2}, for $\|x_t - \hat{x}_{t|t}\| \leq \Bar{s}$, we can make $\mu$ small enough, such that
\begin{equation}\nonumber
\begin{aligned}
\gamma_{\xi} \left( 2\underline{\gamma}_{\xi}^{-1}
\Big(
3\mu \cdot \|{x}_{t} - \hat{x}_{t|t}\|^2 
\Big) \right) \leq \lambda \|{x}_{t} - \hat{x}_{t|t}\|,
\\
\gamma_{\zeta}\left( 2\underline{\gamma}_{\zeta}^{-1}
\Big(
3\mu \cdot \|{x}_{t} - \hat{x}_{t|t}\|^2 
\Big) \right)  \leq \lambda \|{x}_{t} - \hat{x}_{t|t}\|.     
\end{aligned}
\end{equation}
Thus, from \eqref{eq.FIE RAS 1}, we can obtain
\begin{equation}\label{eq.inductive step}
\begin{aligned}
\|\hat{x}_{t+T|t+T} - x_{t+T}\|
&\leq \lambda \|{x}_{t} - \hat{x}_{t|t}\| 
\\
&\oplus \alpha_{\xi}(\| {\xi} \|_{t:t+T-1})
\oplus \alpha_{\zeta}(\| {\zeta} \|_{t+1:t+T}),    
\end{aligned}
\end{equation}
for $\|x_t - \hat{x}_{t|t}\| \leq \Bar{s}$.

By Proposition \ref{proposition.FIE RAS}, for $i\leq T$, we have
\small{\begin{equation}\label{eq.FIE RAS 2}
\begin{aligned}
&\|\hat{x}_{i|i} - x_{i}\|
\\
\leq &\gamma_x\left(
2\sqrt{3} \cdot \|{x}_{0} - \hat{x}_{0|0}\|, i \right) \oplus \gamma_x \left(2\sqrt {\frac{3T}{\mu} \cdot \Bar{\gamma}_{\xi} \left( \| \xi \|_{0:i-1} \right) }, i \right)
\\
\oplus &\gamma_x\left(
2\sqrt{\frac{3 
T}{\mu} \cdot \Bar{\gamma}_{\zeta} \left( \| \zeta \|_{1:i} \right)
}, i \right) 
\oplus
\gamma_{\xi} \left( 2\underline{\gamma}_{\xi}^{-1}
\Big(
3\mu \cdot \|{x}_{0} - \hat{x}_{0|0}\|^2 
\Big) \right)
\\
\oplus&
\gamma_{\xi} \left( 2\underline{\gamma}_{\xi}^{-1}
\Big(
3T \cdot \Bar{\gamma}_{\xi} \left( \| \xi \|_{0:i-1} \right) \Big) \right) \oplus
\gamma_{\xi} \left( 2\underline{\gamma}_{\xi}^{-1}
\Big(
3T \cdot \Bar{\gamma}_{\zeta} \left( \| \zeta \|_{1:i} \right) \Big) \right)
\\
\oplus& \gamma_{\zeta}\left( 2\underline{\gamma}_{\zeta}^{-1}
\Big(
3\mu \cdot \|{x}_{0} - \hat{x}_{0|0}\|^2 
\Big) \right) 
\oplus
\gamma_{\zeta}\left(2\underline{\gamma}_{\zeta}^{-1}
\Big(
3T \cdot \Bar{\gamma}_{\xi} \left( \| \xi \|_{0:i-1} \right) \Big) \right)
\\
\oplus& \gamma_{\zeta}\left(2\underline{\gamma}_{\zeta}^{-1} \Big(
3T \cdot \Bar{\gamma}_{\zeta} \left( \| \zeta \|_{1:i} \right) \Big) \right).
\end{aligned}
\end{equation}}
Regarding the definition of $\alpha_{\xi}, \alpha_{\zeta}$ and define the $\mathcal{K}$ function $\alpha_x$ as 
\begin{equation}\nonumber
\alpha_x(s) = \gamma_x(
2\sqrt{3} s, 0) \oplus \lambda s, 
\end{equation}
we have
\begin{equation}\nonumber
\begin{aligned}
\|x_{i} - \hat{x}_{i|i}\| \leq \alpha_x(\|x_{0} -\hat{x}_{0|0}\|) \oplus \alpha_{\xi}(\| {\xi} \|_{0:i-1})
\oplus \alpha_{\zeta}(\| {\zeta} \|_{1:i}).
\end{aligned}
\end{equation}
Next, we prove by induction that
\begin{equation}\label{eq.induction base}
\begin{aligned}
&\|\hat{x}_{i+jT|i+jT} - x_{i+jT}\| 
 \leq
 \lambda^j \alpha_x(\|x_{0} -\hat{x}_{0|0}\|) 
\\ &\oplus \alpha_{\xi}(\| {\xi} \|_{jT:jT+i-1})
\oplus \alpha_{\zeta}(\| {\zeta} \|_{jT+1:jT+i}) 
\\
& \oplus \max_{0\leq k \leq j-1} \left\{ \alpha_{\xi}(\lambda^{j-k-1} \| {\xi} \|_{kT:(k+1)T-1}) \right\}
\\ &\oplus \max_{0\leq k \leq j-1} \left\{ \alpha_{\zeta}(\lambda^{j-k-1} \| {\zeta} \|_{kT+1:(k+1)T}) \right\},   
\end{aligned}
\end{equation}
for all $j \geq 0$ and $0 \leq i \leq T-1$ with $\lambda < 1$. The base is \eqref{eq.induction base} and then we perform the inductive step.
By applying \eqref{eq.inductive step}, we have
\begin{equation}\nonumber
\begin{aligned}
& \|\hat{x}_{i+(j+1)T|i+(j+1)T} - x_{i+(j+1)T}\| 
\\
\leq 
& \lambda \|{x}_{i+jT} - \hat{x}_{i+jT|i+jT}\| \oplus \alpha_{\xi}(\| {\xi} \|_{i+jT:i+(j+1)T-1})
\\
\oplus & \alpha_{\zeta}(\| {\zeta} \|_{i+jT+1:i+(j+1)T}) 
\\
\leq 
& \lambda^{j+1} \alpha_x(\|x_{0} -\hat{x}_{0|0}\|) \oplus \lambda \alpha_{\xi}( \| {\xi} \|_{jT:jT+i-1})
\\
\oplus & \lambda \alpha_{\zeta}( \| {\zeta} \|_{jT+1:jT+i}) 
\oplus \max_{0\leq k \leq j-1} \left\{ \alpha_{\xi}(\lambda^{j-k} \| {\xi} \|_{kT:(k+1)T-1}) \right\}
\\
\oplus & \max_{0\leq k \leq j-1} \left\{ \alpha_{\zeta}(\lambda^{j-k} \| {\zeta} \|_{kT+1:(k+1)T}) \right\}
\\
\oplus & \alpha_{\xi}(\| {\xi} \|_{i+jT:i+(j+1)T-1})
\oplus \alpha_{\zeta}(\| {\zeta} \|_{i+jT+1:i+(j+1)T}) 
\\
\leq 
& \lambda^{j+1} \alpha_x(\|x_{0} -\hat{x}_{0|0}\|) \oplus \alpha_{\xi}(\| {\xi} \|_{jT:(j+1)T+i-1})
\\
\oplus & \alpha_{\zeta}(\| {\zeta} \|_{jT+1:(j+1)T+i}) 
\oplus \max_{0\leq k \leq j-1} \left\{ \alpha_{\xi}(\lambda^{j-k} \| {\xi} \|_{kT:(k+1)T-1}) \right\}
\\
\oplus & \max_{0\leq k \leq j-1} \left\{ \alpha_{\zeta}(\lambda^{j-k} \| {\zeta} \|_{kT+1:(k+1)T}) \right\}
\\
\leq
&  \lambda^{j+1} \alpha_x(\|x_{0} -\hat{x}_{0|0}\|) \oplus \alpha_{\xi}(\| {\xi} \|_{(j+1)T:(j+1)T+i-1})
\\
\oplus & \alpha_{\zeta}(\| {\zeta} \|_{(j+1)T+1:(j+1)T+i}) 
\oplus  \max_{0\leq k \leq j} \left\{ \alpha_{\xi}(\lambda^{j-k} \| {\xi} \|_{kT:(k+1)T-1}) \right\}
\\
\oplus & \max_{0\leq k \leq j} \left\{ \alpha_{\zeta}(\lambda^{j-k} \| {\zeta} \|_{kT+1:(k+1)T}) \right\}.    
\end{aligned}
\end{equation}
which is the required statement.
An immediate consequence of this statement is that
\begin{equation}\nonumber
\begin{aligned}
&\|\hat{x}_{i+jT|i+jT} - x_{i+jT}\| \\
 \leq &
 \lambda^j \alpha_x(\|x_{0} -\hat{x}_{0|0}\|) \oplus \alpha_{\xi}(\| {\xi} \|_{0:jT+i-1})
\oplus \alpha_{\zeta}(\| {\zeta} \|_{1:jT+i}),
\end{aligned}
\end{equation}
for all $0 \leq i \leq T-1$ and $j \geq 0$. For $t=jT+i$, define the $\mathcal{KL}$ function $\rho_x(s, t) := \lambda^{\lfloor{t/T}\rfloor} \alpha_x(s), \rho_{\xi} = \alpha_{\xi}, \rho_{\zeta} = \alpha_{\zeta}$, thus we have
\begin{equation}\nonumber
|x_t - \hat{x}_{t|t}| \leq \rho_x(\|x_0 - \hat{x}_{0|0} \|, t) \oplus \rho_{\xi}(\| {\xi} \|_{0:t-1})  \oplus \rho_{\zeta}(\| {\zeta} \|_{1:t}).     
\end{equation}
\end{proof}
\end{document}